\documentclass[]{pasj02} 
\usepackage[switch,mathlines]{lineno} 
\renewcommand\makeLineNumber{}

\jyear{2024}
\Received{}
\Accepted{}

\usepackage{romannum}
\usepackage{url}

\newcommand{\be}{\begin{eqnarray}}
\newcommand{\ee}{\end{eqnarray}}

\begin{document}
\nolinenumbers
\title{Testing General Relativity with \textsl{NuSTAR} data of Galactic Black Holes : \Romannum{2}}


\author{
Ashutosh \textsc{Tripathi},\altaffilmark{1,2,3,4}\altemailmark\orcid{0000-0002-3960-5870} \email{ashutosh31tripathi@gmail.com} 
Swarnim \textsc{Shashank}, \altaffilmark{2} \orcid{0000-0003-3402-7212} \email{swarrnim@gmail.com}
Gitika \textsc{Mall}, \altaffilmark{2}
and
Askar~B.\textsc{~Abdikamalov} \altaffilmark{5,2,6}
}

\altaffiltext{1}{Xinjiang Astronomical Observatory, CAS, 150 Science-1 Street, Urumqi 830011, People's Republic of China}

\altaffiltext{2}{Center for Astronomy and Astrophysics, Center for Field Theory and Particle Physics, and Department of Physics,
Fudan University, Shanghai 200438, China}

\altaffiltext{3}{George P. and Cynthia Woods Mitchell Institute for Fundamental Physics and Astronomy, Texas A\&M University, College Station, TX 77843-4242, USA}

\altaffiltext{4}{Department of Physics, Southern Methodist University, 3215 Daniel Avenue, Dallas, Texas 75205, USA} 

\altaffiltext{5}{School of Natural Sciences and Humanities, New Uzbekistan University, Tashkent 100007, Uzbekistan}

\altaffiltext{6}{Ulugh Beg Astronomical Institute, Tashkent 100052, Uzbekistan}



\maketitle

\begin{abstract}
General relativity predicts the spacetime metric around an astrophysical black hole to be described by the Kerr solution, which is a massive rotating black hole without any residual charge. 
In a previous paper, we analyzed the \textsl{NuSTAR} observations of six X-ray binaries to obtain constraints on the deformation parameter $\alpha_{13}$ using a state-of-the-art relativistic model. In this work, we continue analyzing \textsl{NuSTAR} observations of four more X-ray Binaries; two of which, namely Swift~J174540.7--290015 and Swift~J174540.2--290037 are X-ray Transients very close to the supermassive black hole at the center of our galaxy.  The other two sources have complicated absorption, which is accounted for by time- and flux-resolved spectroscopy. The observation of MAXI J1631-479 is divided into two parts to account for the sudden increase in flux. The V404 Cygni spectra, obtained by combining two consecutive observations, are divided into 5 flux states and also account for absorption by quantifying the excess flux in the energy range of 6.5-7.0 keV. The constraints obtained are consistent with the Kerr hypothesis and are comparable with those obtained in previous studies and those from gravitational events. This work shows that even highly absorbed sources can be used for testing the Kerr hypothesis, which is possible with careful data reduction and subsequent data analysis.

\end{abstract}
\pagewiselinenumbers 

\maketitle


\section{Introduction}

In 1915, Albert Einstein proposed the theory of General Relativity (GR)~\citep{Einstein:1916vd}, and after four years, it successfully passed the test conducted by Eddington during the total solar eclipse. After this experiment in a weak field regime, this theory has passed other such tests in the solar system and in observations of binary pulsars~\citep{Will:2014kxa}. After passing successfully in the weak-field limit, there has been an increased interest in the scientific community in testing this theory in a strong-field regime. Thanks to technological advancements, the last decade has seen an immense improvement in
testing this theory, and now we can test the predictions of GR with X-ray data~\citep{Cao:2017kdq, Tripathi:2018lhx, Tripathi:2020qco,2021ApJ...913...79T}, 
Very Long Baseline Interferometry (VLBI)~\citep{Psaltis:2020lvx} and gravitational wave (GW) observations~\citep{TheLIGOScientific:2016src,Yunes:2016jcc,LIGOScientific:2019fpa}.  

Black holes are four-dimensional solutions of the Einstein equations. The uncharged spinning black holes are given by the Kerr solution~\citep{Kerr:1963ud}, which describes the spacetime metric around the astrophysical black holes. This is the direct consequence of ``no-hair” theorems, where the black
hole is simply quantified by three quantities: mass, spin, and charge.  It is also found that deviations caused by non-vanishing electric charge, nearby stars, and the accretion disk are negligible~\citep{2014PhRvD..89l7302B,2018AnP...53000430B}. There are several scenarios in which
macroscopic deviations from the Kerr solution are possible~\citep{Giddings:2017jts, Herdeiro:2014goa}. These scenarios include models with macroscopic quantum gravity effects or models with the presence of exotic matter fields. It is also possible that there is a classical extension of GR and the current form of GR is not the correct theory of gravity~\citep{Kleihaus:2011tg}. Thus, testing the Kerr hypothesis around the black holes is a means by which GR in a strong-field regime can be tested.

X-ray Reflection Spectroscopy is one of the most suitable methods to determine the properties of black holes by studying the reflection spectrum emitted by the accretion disk around it~\citep{2006ApJ...652.1028B}. The schematic representation of an astrophysical black hole system is shown in Fig.~\ref{fig:1}. For stellar-mass black holes, the temperature of the accretion disk lies in the soft X-rays (0.1-1.0 keV). A part of the thermal component interacts with the electrons via the inverse Compton scattering present in the corona, which is a hot (~100 keV) and optically thin medium above the black hole. This inverse Compton scattering generates a power-law component with a cutoff energy of about 300 keV. A part of this power-law component illuminates the disk and produces the reflection spectrum, which constitutes various emission and absorption features. The most prominent reflection signatures are the iron K$\alpha$ line around 6 keV and the Compton hump around 20 keV\citep{Ross:2005dm, Garcia:2010iz}. The reflection spectrum at any point on the disk in the rest frame of the gas is determined by atomic physics, where the atomic transitions of ionized ions are treated in detail. The photons travel in the strong gravitational field present around black holes and are affected by relativistic effects such as light bending, Doppler broadening, gravitational redshift, etc., before reaching the observer. In the presence of high-resolution data (so that the reflection features can be resolved) and the correct astrophysical effects (to account for the relativistic effects and emission lines), we can thus probe the regions that are very close to the black hole~\citep{2019NatAs...3...41R}. We can also measure the reflection spectrum accurately, which will be helpful in determining the space-time metric around the black hole and thus test the Kerr hypothesis around astrophysical black holes. 

Our group has developed a model {\tt relxill\_nk} \citep{Bambi:2016sac, Abdikamalov:2019yrr} to test the Kerr hypothesis using X-ray Reflection Spectroscopy around black holes. {\tt relxill\_nk} is the extension of {\tt relxill} \citep{Dauser:2013xv, Garcia:2013lxa} to non-Kerr metrics such as Johannsen metric \citep{2013PhRvD..88d4002J}, KRZ metric \citep{2016PhRvD..93f4015K}, etc. The reflection spectrum at an emission point in the rest frame of the gas in the accretion disk is modified by the relativistic convolution model for a background metric. The deviations from the Kerr solutions in the background metric are quantified by ``deformation parameters” which vanish for the Kerr solution. Comparison of the X-ray observations of black holes against the theoretical modeling of {\tt relxill\_nk} measures the deformation parameters and thus tests the Kerr hypothesis in a strong-field regime. 

In the past five years, we employed {\tt relxill\_nk} to constrain the deformation parameters using X-ray observations of various X-ray binaries and supermassive black holes~\citep{Bambi:2015kza, Bambi:2016sac, 2021SSRv..217...65B, Abdikamalov:2019yrr, Abdikamalov:2020oci}. The most stringent constraint obtained is for the simultaneous \textsl{Swift} and \textsl{NuSTAR} observation of GX 339-4~\citep{Tripathi:2020dni}. 
All measurements of deformation parameters are consistent with the Kerr hypothesis. It is important to note that similar tests of Kerr hypothesis have also been performed with GW observations \citep{Cardenas-Avendano:2020xtw, 2022PhRvD.105j4004S} and Event Horizon Telescope observations \citep{Psaltis:2020lvx, 2022ApJ...930L..17E}.

In~\citet{2021ApJ...913...79T}, we applied {\tt relxill\_nk} to analyze the reflection spectrum of six X-ray binaries observed by \textsl{NuSTAR} observations. We obtained the most robust constraints on the Kerr hypothesis available to date. The constraints of these X-ray binaries are comparable to those obtained from gravitational waves and other electromagnetic techniques. Motivated by our results, we selected complicated sources from the list of all published spin measurements using \textsl{NuSTAR}. For these observations, the inner edge of the accretion is very close to the black hole, which is considered favorable for obtaining tight constraints on the deformation parameters. Also, these sources are found to have high spin, which is also an essential requirement for getting stringent constraints on the deviations from the Kerr hypothesis. 

This manuscript is organized as follows. In Section~\ref{s:obs}, we will briefly describe the selection of sources and observations. The spectral analysis for each source is presented in Section ~\ref{s:ana}. In Section ~\ref{s:dis}, we discuss and summarize the results. Throughout the paper, we adopt the convention (-+++) and $G_N$=c=1.

\begin{figure}[t]
\begin{center}
\includegraphics[width=8.5cm,trim={0cm 0cm 0cm 0cm},clip]{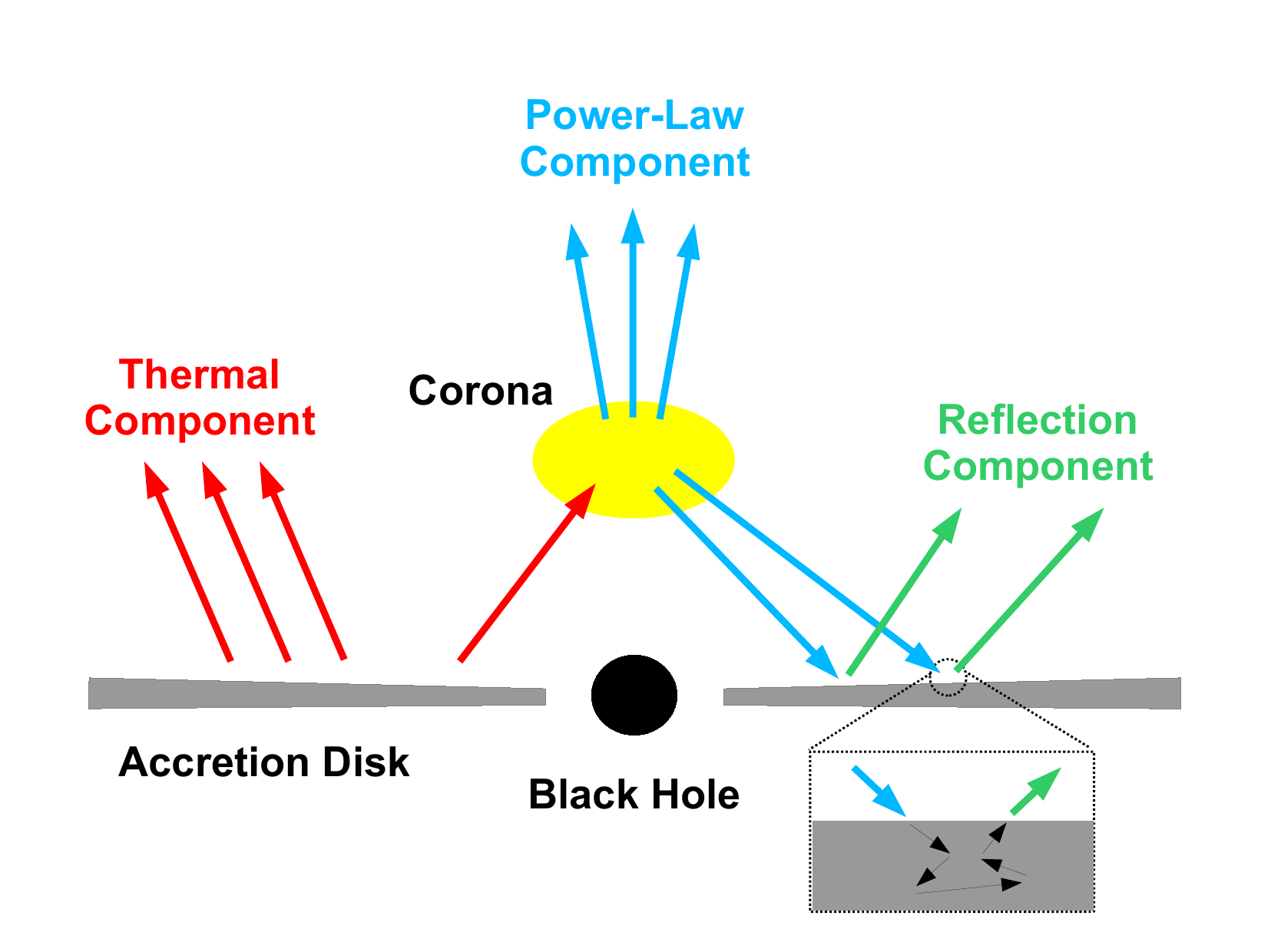}
\end{center}
\caption{A black hole is accreting from a geometrically thin and optically thick disk. The disk has a multi-temperature blackbody-like spectrum. Thermal photons from the disk inverse Compton scatter off free electrons in the corona, producing the continuum component. The latter illuminates the disk, generating the reflection component. From \citet{2021ApJ...913...79T}\\
ALT text : Corona.}
\end{figure}\label{fig:1}


\section{Selection of the sources and data reduction} \label{s:obs}

We started with the list of all published spin measurements done by \textsl{NuSTAR}~\citep{Harrison:2013md}. \textsl{NuSTAR} is considered to be currently the best X-ray mission to study the reflection features originating in the innermost regions of black holes, where the effects of gravity are immense. The main advantage of \textsl{NuSTAR} is the coverage of a wide energy range of 3-79~keV, which includes two of the most prominent reflection features: iron line and Compton hump. Besides, pile-up is absent, which is usually the dominant effect for the bright sources. 


In ~\citep{2021ApJ...913...79T}, we selected six sources from the list of fourteen objects given in Tab.~\ref{t-bhb} that have a reflection spectrum minimally affected by absorption, dust scattering, and have an accretion disk that extends up to ISCO. These six sources are GX 339-4, Swift~J168.2—4242, 4U~1630–472, GRS~1739–278, GS~1354-645, and EXO~1846–031. In this paper, we analyzed the observations for which the variability and absorptions could be modeled by applying advanced spectroscopic and reduction methods like flux-resolved, flare-resolved, and time-resolved spectroscopy, etc. In this work, we analyze the \textsl{NuSTAR} observations of four sources:  Swift~J174540.7--290015, Swift~J174540.2--290037, MAXI~J1631--479, and V404 Cygni. 

The sources are highly absorbed sources and already had spin measurement using observations analyzed in this paper (see reference in Table 1 for details). These objects meet both criteria for testing the Kerr hypothesis: high spins and prominent reflection signatures. If we have a spin measurement with these highly absorbed sources, then it is also possible to test the Kerr hypothesis. In \citet{2014ApJ...787...83M}, $\approx350$ ks of simultaneous \textsl{XMM-Newton} and \textsl{NuSTAR} observations of Narrow line Seyfert 1 galaxy MCG-06-3015 are analyzed. This source shows prominent reflection features in this observation, but is highly variable due to the presence of some warm absorbers and dust around it, which is also confirmed with high-resolution Chandra spectroscopy (see \citet{2014ApJ...787...83M} and references therein). The highly variable nature is  quantified by time-resolved spectroscopy, where the spectra are obtained for different time intervals for which the flux variability is not significant, and then are analyzed simultaneously to obtain the spin measurement. This analysis provides very stringent constraints on spin and other key parameters. Thus, this study shows that the variability of the source is accounted for by careful analysis. In this paper, a similar kind of approach is used to properly assess the variability before constraining the spin and, consequently, the deformation parameter. With these 4 sources, we aim to demonstrate that highly variable data can also be used for such a study, provided the data is reduced and analyzed appropriately.

\begin{table*}
    \centering
    \renewcommand\arraystretch{1.5}{
    \begin{tabular}{ccccc}
    \hline\hline
    Source & $a_*$ & Mission(s) & State & Reference \\
    \hline\hline
    4U~1630--472 & $0.985_{-0.014}^{+0.005}$ & \textsl{NuSTAR} & Intermediate & \citet{King:2014sja} \\
    \hline    
    Cyg~X-1& $>0.83$ & \textsl{NuSTAR}+\textsl{Suzaku} & Soft & \citet{Tomsick:2013nua} \\
    & $>0.97$ & \textsl{NuSTAR}+\textsl{Suzaku} & Hard & \citet{Parker:2015fja} \\
    & $0.93 \sim 0.96$ & \textsl{NuSTAR} & Soft & \citet{Walton:2016hvd} \\
    \hline
    EXO 1846--031 & $0.997_{-0.002}^{+0.001}$ & \textsl{NuSTAR} & Hard intermediate & \citet{Draghis:2020ukh} \\
    \hline    
    GRS~1716--249 & $>0.92$ & \textsl{NuSTAR}+\textsl{Swift} & Hard Intermediate & \citet{Tao:2019yhu} \\
    \hline
    GRS~1739--278 & $0.8 \pm 0.2$ & \textsl{NuSTAR} & Low/Hard & \citet{Miller:2014sla} \\
    \hline
    GRS~1915+105 & $0.98 \pm 0.01$ & \textsl{NuSTAR} & Low/Hard & \citet{Miller:2013rca} \\
    \hline    
    GS~1354--645 & $>0.98$ & \textsl{NuSTAR} & Hard & \citet{El-Batal:2016wmk} \\
    \hline
    GX~339--4 & $0.95_{-0.08}^{+0.02}$ & \textsl{NuSTAR}+\textsl{Swift} & Very High & \citet{Parker:2016ltr} \\
    \hline
    MAXI~J1535--571 & $>0.84$ & \textsl{NuSTAR} & Hard & \citet{Xu:2017yrm} \\
    \hline
    MAXI~J1631--479 & $>0.94$ & \textsl{NuSTAR} & Soft & \citet{Xu:2020vil} \\
    \hline    
    Swift~J1658.2--4242 & $>0.96$ & \textsl{NuSTAR}+\textsl{Swift} & Hard & \citet{Xu:2018lfo} \\
    \hline
    Swift~J174540.2--290037 & $0.92_{-0.07}^{+0.05}$ & \textsl{Chandra}+\textsl{NuSTAR} & Hard & \citet{Mori:2019iwz} \\    
    \hline
    Swift~J174540.7--290015 & $0.94_{-0.10}^{+0.03}$ & \textsl{Chandra}+\textsl{NuSTAR} & Soft & \citet{Mori:2019iwz} \\
    \hline
    V404~Cyg & $>0.92$ & \textsl{NuSTAR} & Hard & \citet{Walton:2016fso}\\
    \hline\hline
\end{tabular} }
\vspace{0.2cm}
\caption{\rm Spin measurements of stellar-mass black holes from the analysis of relativistic reflection features of the sources using \textsl{NuSTAR} observations. From \citet{2021ApJ...913...79T}.}
\label{t-bhb}
\end{table*}


\subsection{Data reduction} \label{s:ana}

\begin{table*}
 \centering
 \renewcommand\arraystretch{1.5}
\begin{tabular}{ccc}
\hline\hline
\hspace{0.1cm} Source \hspace{0.1cm} & \hspace{0.1cm} Observation ID \hspace{0.1cm} & \hspace{0.1cm} Observation Date \hspace{0.1cm}\\
\hline\hline
SWIFT J174540.7-290015              & 90101022002 & 2016 February 22   \\
\hline
SWIFT J174540.2-290037             & 90201026002 & 2016 June 9  \\
\hline
MAXI J1631-479             & 90501301001& 2019 Jan 17  \\

\hline
V404Cygni            & 90102007002/3 & 2015 July 11  \\\hline\hline
\end{tabular}
 \caption{\rm Summary of the sources and the observations analyzed in the present work. \label{t-obs}}
\end{table*}

Tab.~\ref{t-obs} presents the details of the observations of four sources analyzed in this work. The unfiltered events from the FPMA and FPMB instruments of \textsl{NuSTAR} are
converted into cleaned events using the latest calibration files and the \textsl{NuSTAR} data analysis software (NuSTARDAS) v2.0.0, which is distributed as part of the HEASOFT package. We select the source region with a center at the source such that around 90 percent of source photons will be captured. The background region is selected as far as possible from the source on the same detector so that the effect of the source is not significant. The end products (the spectra and the light curves) are generated using the {\tt nupipeline} routine of NuSTARDAS. The cross-calibration constant between FPMA is kept frozen at 1.0 and thawed for the FPMB of each observation. We will discuss the details of the data reduction for each source separately.

\section{Spectral analysis}
\label{s:ana}
Various models are employed to explain the different emissions from black hole systems. Now, we will discuss these models. 

{\tt Tbabs} \citep{2000ApJ...542..914W} account for X-ray absorption in the interstellar medium (ISM). Here, the free parameters are the hydrogen equivalent column density. The abundance of other elements is assumed to be the default in this work. In some sources with high absorption, we use {\tt Tbnew} \citep{2000ApJ...542..914W}, which allows us to change the column density of other elements and takes into account the depletion of elements. We used {\tt Tbnew} in all the sources except for V404, where the variability is quantified by flux-resolved spectroscopy. For simplicity, we kept all parameters frozen except the column density of hydrogen.

The accretion disk around the black hole is assumed to be optically thick and geometrically thin. The temperature of the disk lies in the soft X-ray band ($\ sim$1 keV) and emits blackbody radiation. The whole accretion disk emits a multi-temperature blackbody-like spectrum. This spectrum is described by the model {\tt diskbb} \citep{1984PASJ...36..741M} in XSPEC. The parameters are the inner temperature of the disk (in keV) and normalization. These thermal photons interact with the electrons present in the corona via Compton inverse scattering. Corona is believed to be a hotter cloud of gas with a temperature of around 100 keV situated above the black hole and the accretion disk. This interaction leads to the generation of a continuum with an exponential energy cut-off. The power-law photon continuum is described by the additive model {\tt powerlaw}. The parameters are photon index $\Gamma$, cut-off energy $E_{cut}$, and normalization. Another model {\tt compTT} \citep{1994ApJ...434..570T} describes the Comptonized continuum and also includes relativistic effects. 

A part of the continuum produced in the corona reaches the observer, and another part illuminates the disk, producing various absorption and emission lines together referred to as the reflection spectrum. The most prominent reflection signatures are iron K$\alpha$ emission around 6-7 keV and the Compton hump around 20 keV. {\tt xillver} \citep{Garcia:2010iz, Garcia:2013lxa} describes the reflection spectrum in the rest frame of the gas, which is not significantly affected by the strong gravity of a black hole. The parameters of this model are the photon index of the incident continuum $\Gamma$, ionization of the disk log$\xi$, iron abundance in solar units $A_{Fe}$, and the reflection fraction $Ref_{frac}$. The reflection spectrum generated in the inner regions of a black hole is affected by its strong gravity. {\tt relxill\_nk} describes the relativistic reflection model assuming that the background metric is a solution given by various non-Kerr metrics. In addition to the parameters of {\tt xillver}, {\tt relxill\_nk} has an emissivity profile, the spin of the black hole $a_*$, and the inclination of the disk $i$ as free parameters. For {\tt xillverCp}, the incident spectrum is assumed to be a Comptonized continuum, and the resulting relativistic reflection model is {\tt relxillCp\_nk}. In {\tt relxillCp\_nk}, the incident spectrum is modeled by {\tt nthComp} and has three key parameters: photon index, electron temperature, and seed photon temperature.

\begin{figure*}
\begin{center}
\includegraphics[scale=0.5]{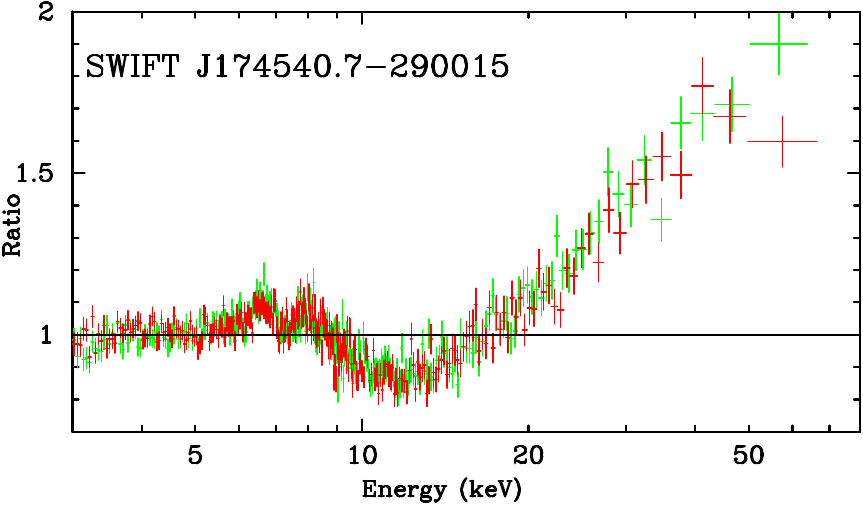}
\includegraphics[scale=0.5]{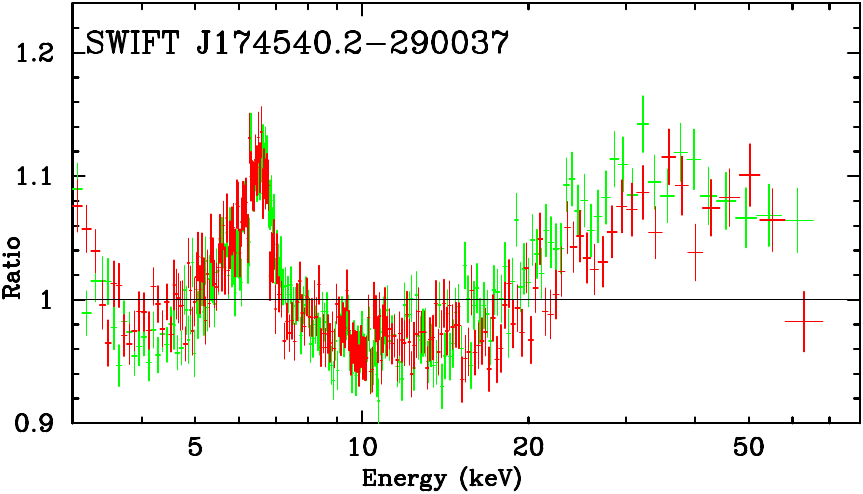} \\
\includegraphics[scale=0.5]{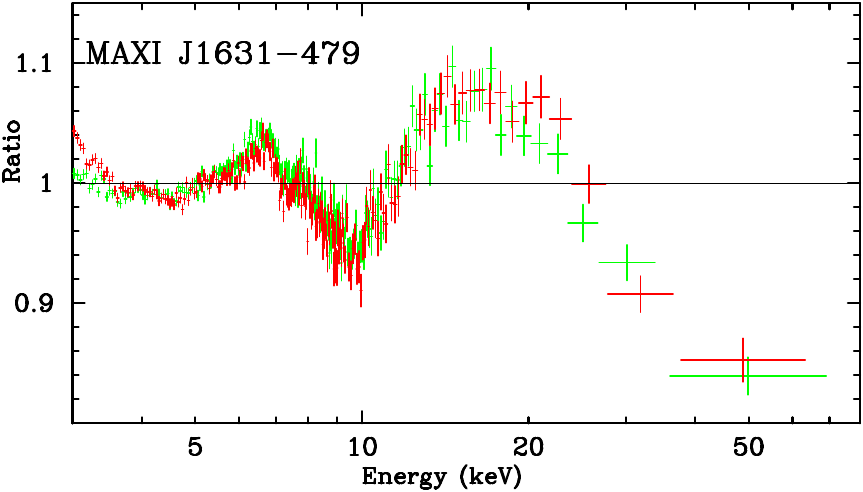}
\includegraphics[scale=0.5]{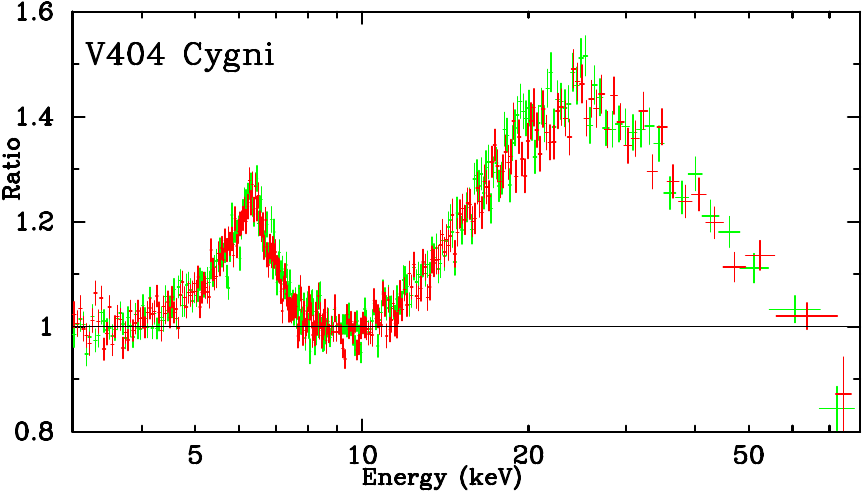} \\

\end{center}
\caption{Data to the best-fit model ratio for an absorbed power-law (an absorbed power-law + disk blackbody spectrum in the case of MAXI J1631-479) for the spectra analyzed in this work. Green and red crosses are for data from FPMA and FPMB instruments of \textsl{NuSTAR}, respectively.\\
ALT text: power-law fit.}\label{fig:abspow}
\end{figure*}

In a standard version of the model, emissivity is assumed to have a broken power law profile described by the inner emissivity index $q_1$, outer emissivity index $q_2$, and the break radius $R_{br}$, where the index changes. The inner radius of the disk is assumed to be at the innermost stable circular orbit (ISCO), and the outer radius is fixed at 400 gravitational radii ($R_g$). When {\tt relxill\_nk} and {\tt xillver} are used to describe a system, the parameters present in both models are tied to each other except the ionization parameter. log$\xi$ in {\tt xillver} is frozen to 0 and that in {\tt relxill\_nk} is kept free to vary. $\Gamma$ and $E_{cut}$ of {\tt relxill\_nk} is tied to corresponding parameters of the continuum. 



We used metallic abundances from \citet{2000ApJ...542..914W} and cross sections from \citet{1996ApJ...465..487V}. XSPEC v12.11.1 is used for the preliminary spectral analysis, for identifying the reflection and absorption features, and for fitting those features to get the best-fit model to the data. We choose the best-fit model which yields the minimum $\chi^2$ statistics. Since we have several parameters, it is possible to have degeneracy among these parameters. To resolve this issue, we run Markov chain Monte Carlo (MCMC) simulations with the priors estimated by the preliminary $\chi^2$ analysis in XSPEC. We use the publicly available package \texttt{xspec\_emcee}\footnote{\url{https://github.com/jeremysanders/xspec_emcee}} to perform our MCMC analysis. We sampled the posterior distribution using 12,000 iterations and 500 random walkers. The errors estimated in this work have 1$\sigma$ confidence, unless stated otherwise. 

Next, we will discuss the data reduction technique and the data analysis method for each of the sources and quote the measurements of spin $a_*$ and $\alpha_{13}$. 
separately. 

\subsection{SWIFT J174540.7-290015}

SWIFT J174540.7-290015 (T15 hereafter) is an X-ray transient discovered north of Sagittarius A$^*$ by \textsl{Neil Gehrels Swift Observatory} on February 6, 2016, and confirmed by \textsl{Chandra} observation taken on 13-14 February 2015, a week after its discovery. T15 was observed by various missions throughout the entire electromagnetic waveband, such as \textsl{XMM-Newton}, \textsl{INTEGRAL/IBIS}, and Very Large Array (VLA). This source was observed for 34 ks by \textsl{NuSTAR} after 16 days of its first detection by \textsl{Swift}, i.e., on February 22, 2016.

The source region is chosen to be a circular region of 30’’ around the source. For the background region, ~\citet{Mori:2019iwz}  performed a detailed procedure to assess the effect of other nearby sources present in the background. They found that the background spectra are negligible ($\sim$2\%)as compared to the source spectra. Hence, we choose a 30’’ circular region on the same detector as the background. The spectra are binned to 30 counts per bin to apply $\chi^2$ statistics.  The extracted spectra have a mean count rate of 6.29 cts$^{-1}$ (FPMA). More details can be found in ~\citet{Mori:2019iwz}.

First, we fit the spectra with {\tt tbnew*powerlaw} to observe the residual features present in the data. Fig.~\ref{fig:abspow} shows the data-to-model ratio for the fit of the spectra with the absorbed power law. The emission around 6-7 keV is visible along with the Compton hump, which peaks around 50 keV. To fit the absorption and reflection features, we employed the following model (in XSPEC jargon):

\begin{equation}
{\tt tbnew*(relxillCp\_nk)}
\end{equation}

Fig.~\ref{ratio} shows the ratio plot for the data fitted to the best-fit model. The reduced chi-square is found to be 1.12. Large residuals are present after 40 keV, indicating that the spectrum is not fitted properly with the reflection models. Following ~\citet{Mori:2019iwz}, we froze the iron abundance at the solar value. If the iron abundance is thawed, it gets pegged to an unreasonably high value, and the spin is loosely constrained. Here, we assumed the emissivity profile as shown by the majority of X-ray binaries \citep{2021ApJ...913...79T}, where the inner emissivity index is found to be very steep, and after the break radius, the outer emissivity nearly follows the Newtonian profile with the value of $\approx$3. Table~\ref{tab-fit} shows the errors associated with the free parameters. The high absorption nature of this source is reflected in the high $n_H$ value of {\tt Tbnew}. The high values of spin and the inclination are found to be consistent with ~\citet{Mori:2019iwz}. The reflection fraction value of less than 1 implies that the spectrum is power law dominated. $\alpha_{13}$ includes the Kerr solution ($\alpha_{13}$ = 0) within 99.73\% (3$\sigma$) percentage confidence range. Fig.~\ref{fig:mcmct15t37} shows the triangle plot for the posterior distribution of the parameters sampled using MCMC analysis. The errors associated with the spin $a_*$ and the deformation parameters are given by :
\begin{equation}
a_* = 0.997^{+0.001}_{-0.001}, \hspace{1cm} \alpha_{13} = -0.016^{+0.032}_{-0.109}
\end{equation}
We also tried to fit the same observation for the Kerr metric ($\alpha_{13}$ = 0). The $\chi^2$ improves by less than 1 when using the non-Kerr model, which is also evident from the value of $\alpha_{13}$ being close to zero.

\subsection{SWIFT J174540.2-290037}

SWIFT J174540.2-290037 (T37 hereafter) was discovered by \textsl{Neil Gehrels Swift Observatory} on May 28, 2016, when T15 was still in outburst. The X-ray transient is situated south of the Sagittarius A$^*$, which is confirmed by its \textsl{Chandra} observations and remained bright for a month. It was observed by \textsl{NuSTAR} for 49 ks 16 days after the onset of the outburst on June 16, 2016 {~\citep{Mori:2019iwz}. 

The source region of 30’’ is taken with the source as the center. For the background spectra, ~\citet{Mori:2019iwz} checked for contamination from other nearby sources and found it to be around 3$\%$, and therefore we neglect the contamination of these sources in extracting the background region. We choose the background region to be of the size of 30" and as far as possible from the source on the same detector. Finally, a binning scheme of 30 counts per bin is employed to make chi statistics applicable following ~\citet{Mori:2019iwz}.  The resultant spectra of the source have an average count rate of 13.6 cts$^{-1}$ (FPMA).

We started the analysis by fitting the spectra with the absorbed power law to highlight the features present in the observation. From Fig.~\ref{fig:abspow}, it is clear that the observations consist of an iron line, a Compton hump, and significant absorption at lower energies. We used the following model combination, similar to that of T15, to explain the black hole system :
\begin{equation}
{\tt tbnew*(relxillCp\_nk)}
\end{equation}

Fig.~\ref{ratio} shows the data fitted to the best-fit model. The upper panel of Fig.~\ref{ratio} shows the reflection model. Tab.~\ref{tab-fit} shows the errors for the parameters of the best-fit model. The reduced $\chi^2$ is found to be 1.06. This system was also found to have a high spin value, as found in the case of T15. However, the inclination angle is lower, and the data favor the broken power-law emissivity profile. The inner emissivity index, like most binary systems, is found to be very steep ($\sim$10). We, therefore, freeze it to 10.0 and fit the outer emissivity and the breaking radius. The outer emissivity is found to be similar to the Newtonian profile, which is also the case for T15. The iron abundance is found to be very high, which is consistent with the results of ~\citet{Mori:2019iwz}. The iron abundance pegs to the highest value of 10 when thawed during the fit and, therefore, is frozen to it. The reflection fraction is found to be 0.37, suggesting that the emission towards the observer is more than that reflected to the disk. The deformation parameter includes the Kerr solution within 3$\sigma$ significance. Fig.~\ref{fig:mcmct15t37} shows the posterior distribution of key parameters obtained by MCMC analysis. The measurements of $a_*$ and $\alpha_{13}$ are given by
\begin{equation}
 a_* = 0.9973^{}_{-0.012}, \hspace{1cm} \alpha_{13} = -0.212^{+0.077}_{-0.095}
\end{equation}
The chi-improves by $\sim$ 6 when the non-Kerr model is used compared to the Kerr model where the value of $\alpha_{13}$ vanishes to zero. 
\begin{figure*}
\begin{center}
\includegraphics[scale=0.3]{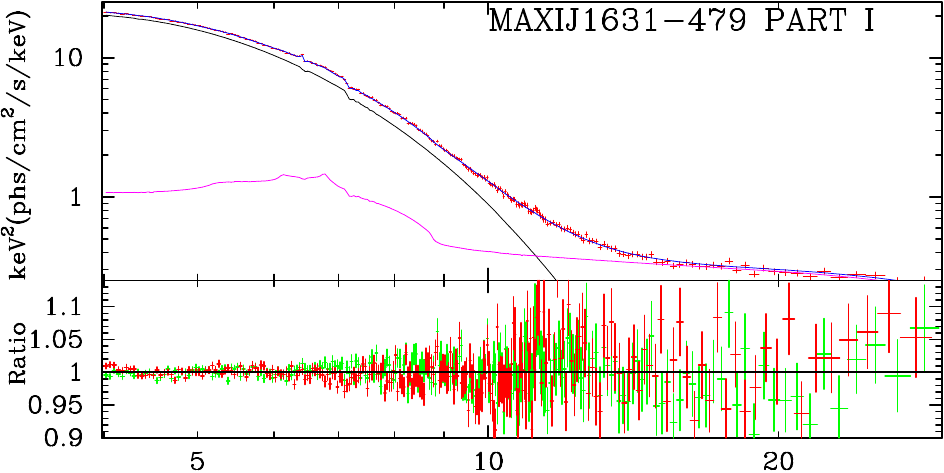}
\includegraphics[scale=0.3]{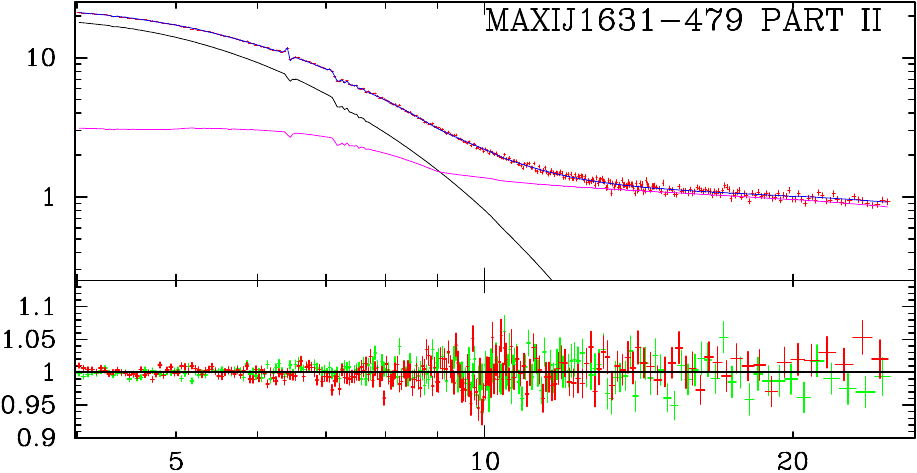} 
\includegraphics[scale=0.3]{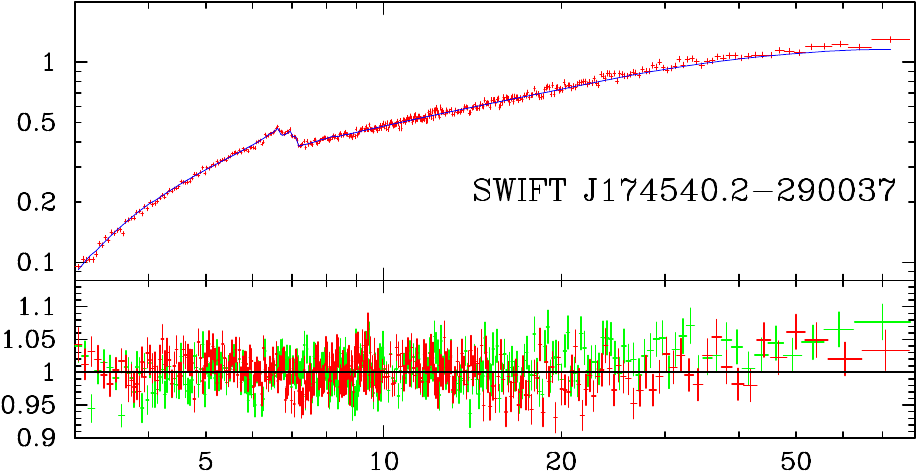} \\
\includegraphics[scale=0.3]{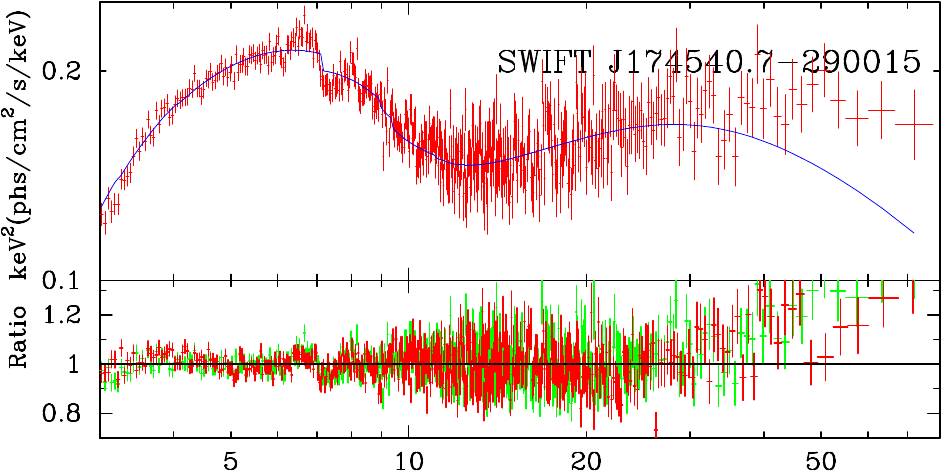}
\includegraphics[scale=0.3]{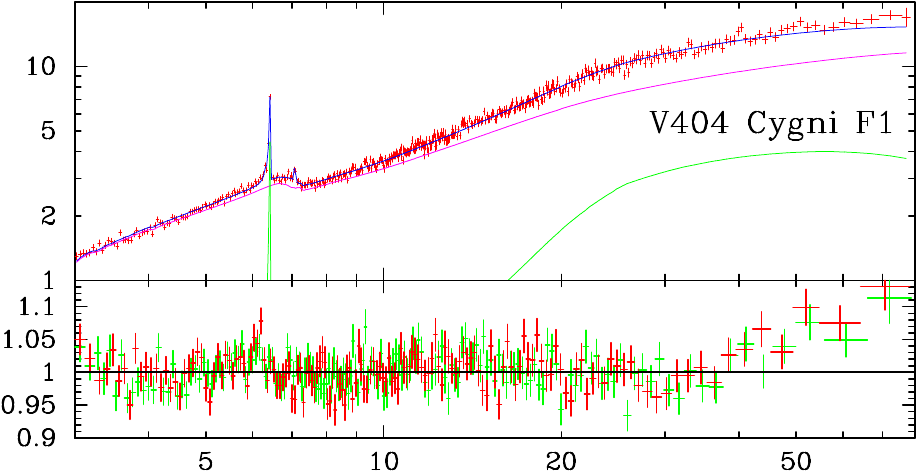} 
\includegraphics[scale=0.3]{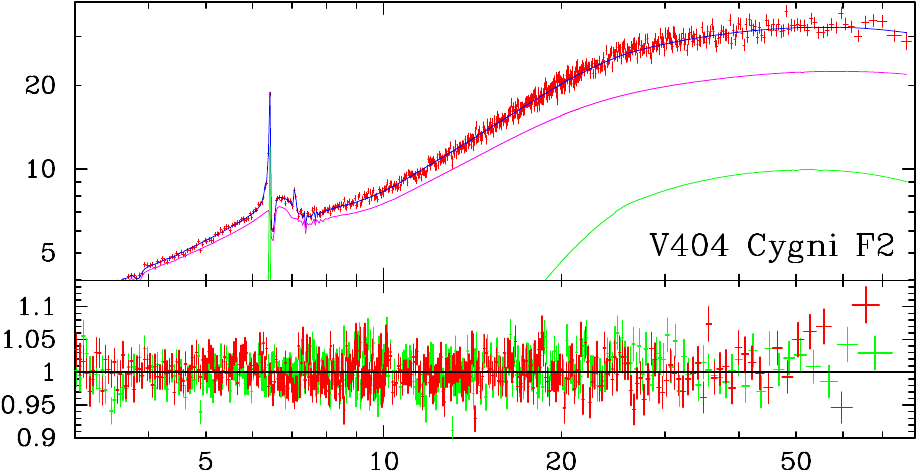} \\
\includegraphics[scale=0.3]{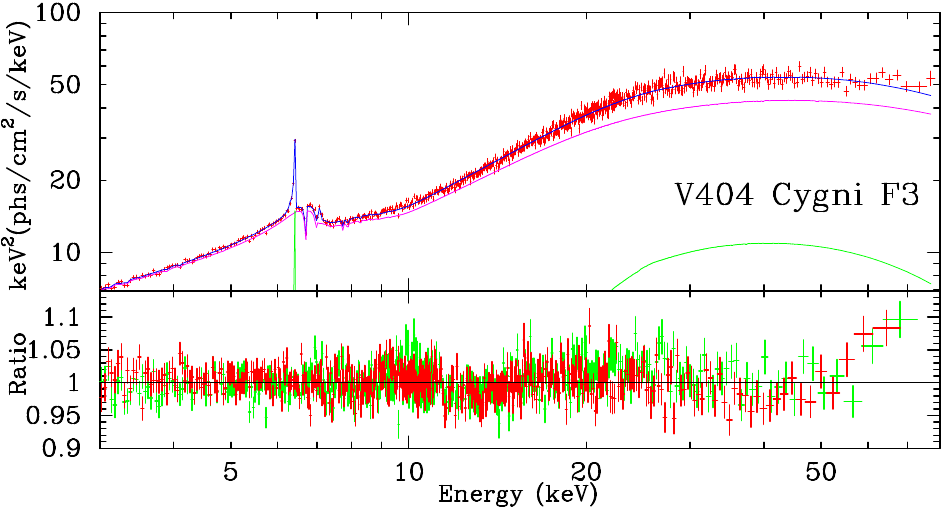}
\includegraphics[scale=0.3]{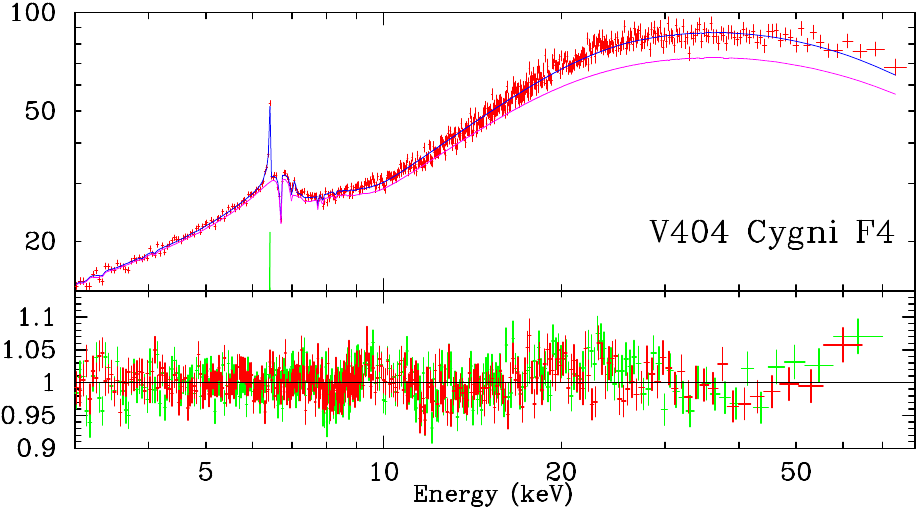} 
\includegraphics[scale=0.3]{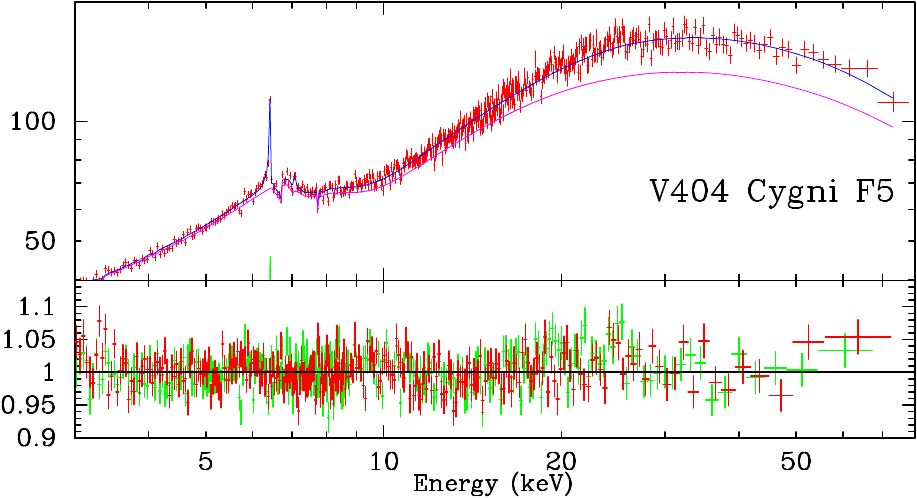} \\
\end{center}
\caption{The best fit model (upper panel) and the ratio of data to the best fit model (lower panel) for various observations analyzed in this work. In the upper panel, the blue curve corresponds to the total model, and the red points correspond to FPMA data (we only show one data for visual clarity). black, purple, and green curves correspond to disk emission, relativistic reflection ((\tt relxill\_nk)), and non-relativistic reflection ((\tt xillver)), respectively. In the lower panels, the green and red curves denote FPMA and FPMB data of \textsl{NuSTAR}, respectively.\\
ALT text: best fit plots.\label{ratio}}
\end{figure*}
\begin{figure*}
\begin{center}
\includegraphics[scale=0.4]{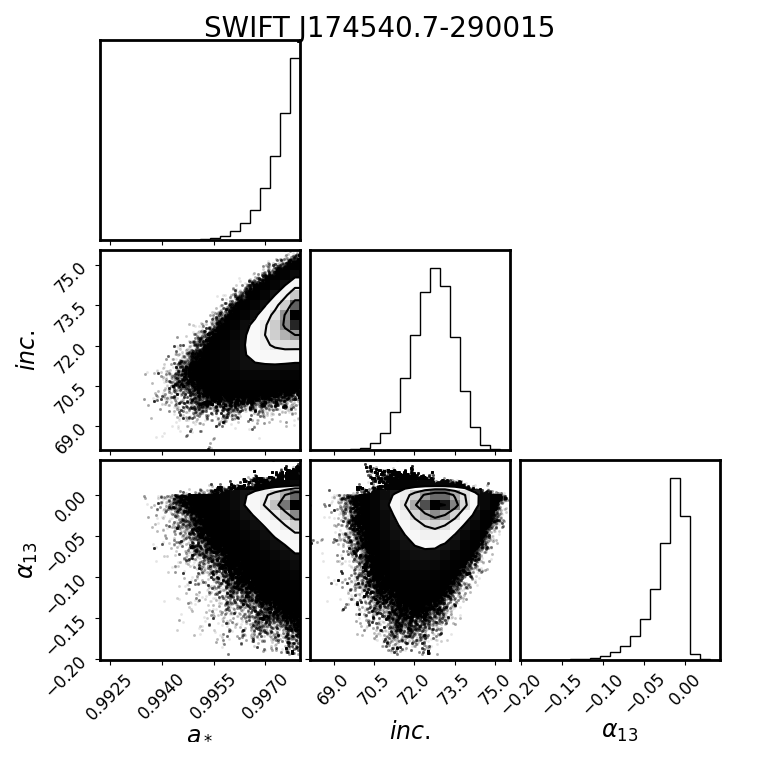}\includegraphics[scale=0.4]{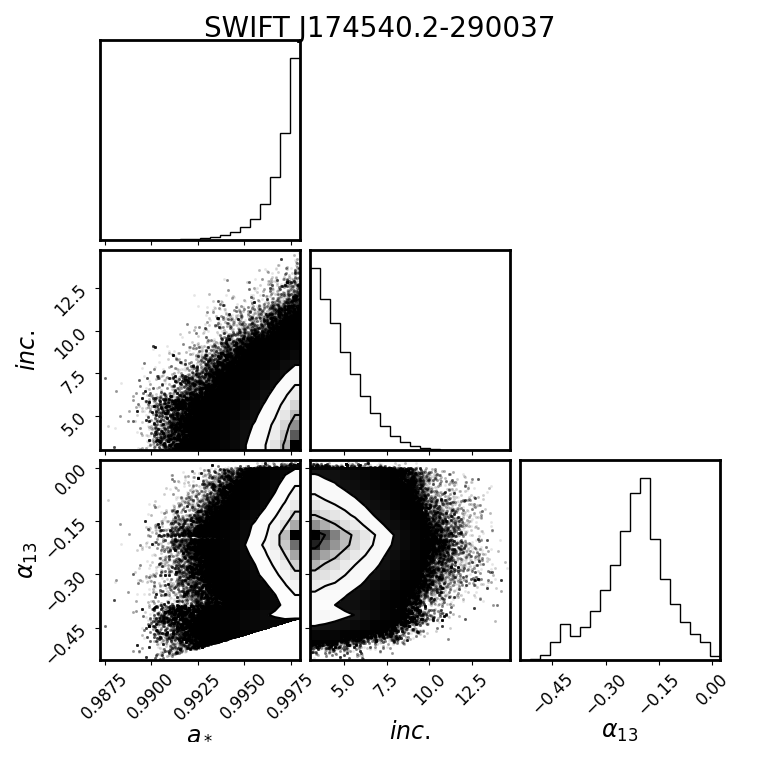}
\end{center}
\vspace{-0.5cm}
\caption{Corner plot for the key parameter-pairs (excluding calibration constants and component normalization) for SWIFT J174540.7-290015 and SWIFT J174540.2-290037 after the MCMC run. The 2D plots report the 1-, 2-, and 3-$\sigma$ confidence contours.\\
ALT text : MCMC plot for T15 and T37.\label{fig:mcmct15t37}}
\end{figure*}

\begin{figure*}
\begin{center}
\includegraphics[scale=0.8]{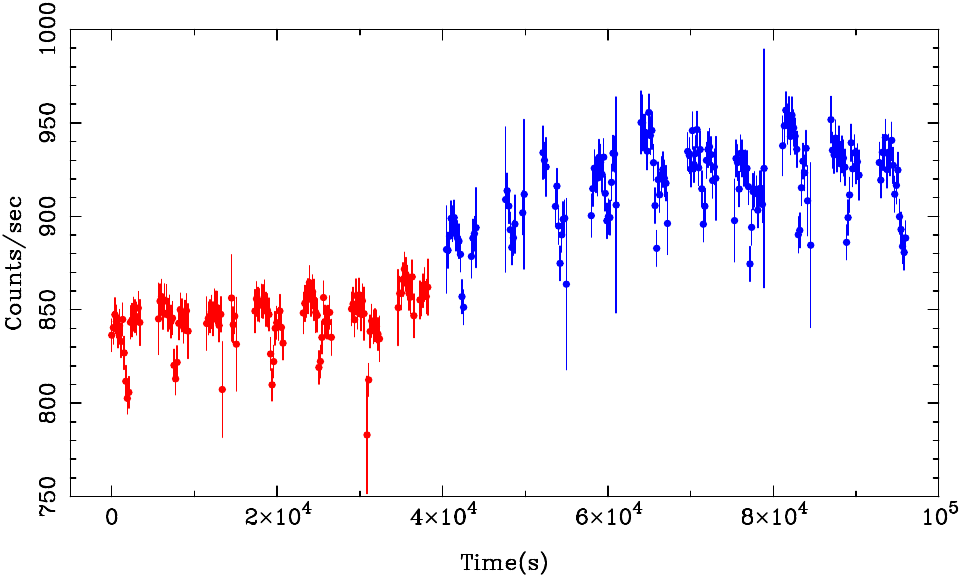}\\
\end{center}
\caption{The light curve of the \textsl{NuSTAR} (FPMB) observation of MAXI J1631--479 for the energy range 3.0-79.0 keV. The red and blue curves correspond to part I and part II of the observation, respectively. \\
ALT text : Light curves for MAXI J1631.}\label{fig:maxi3}
\end{figure*}

\begin{figure*}
\begin{center}
\hspace{-2cm}\includegraphics[scale=0.4]{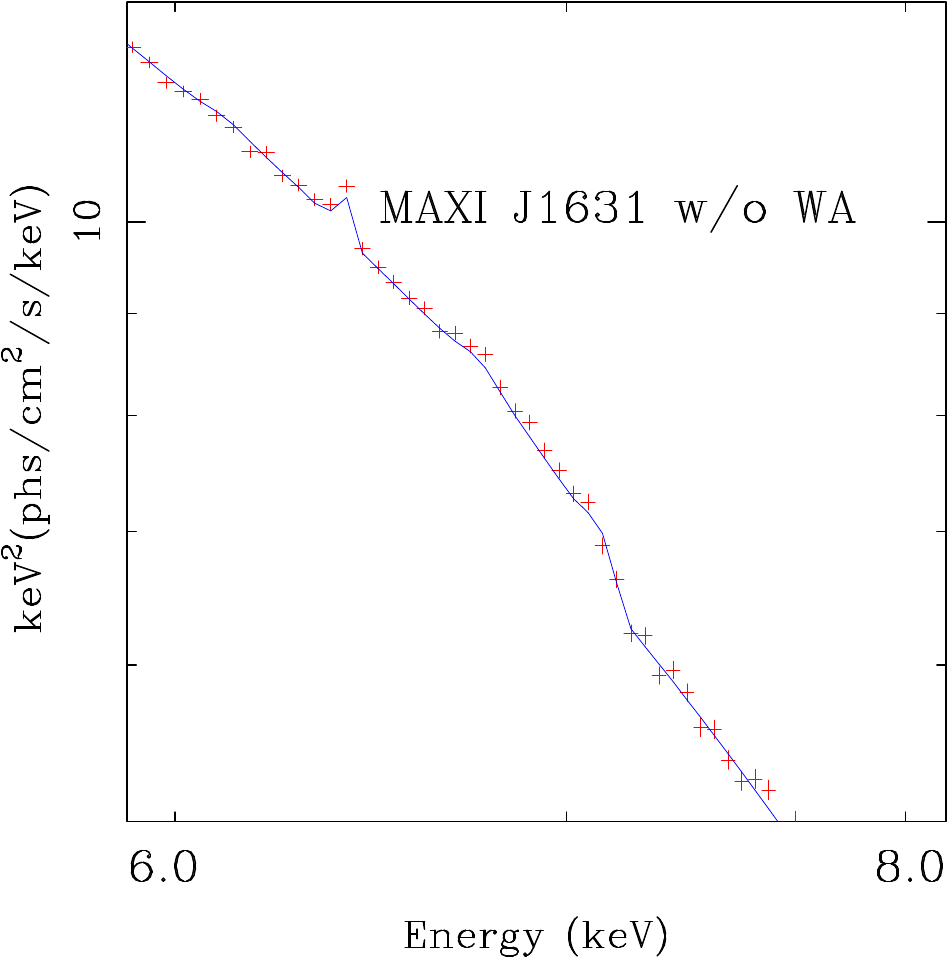}\includegraphics[scale=0.4]{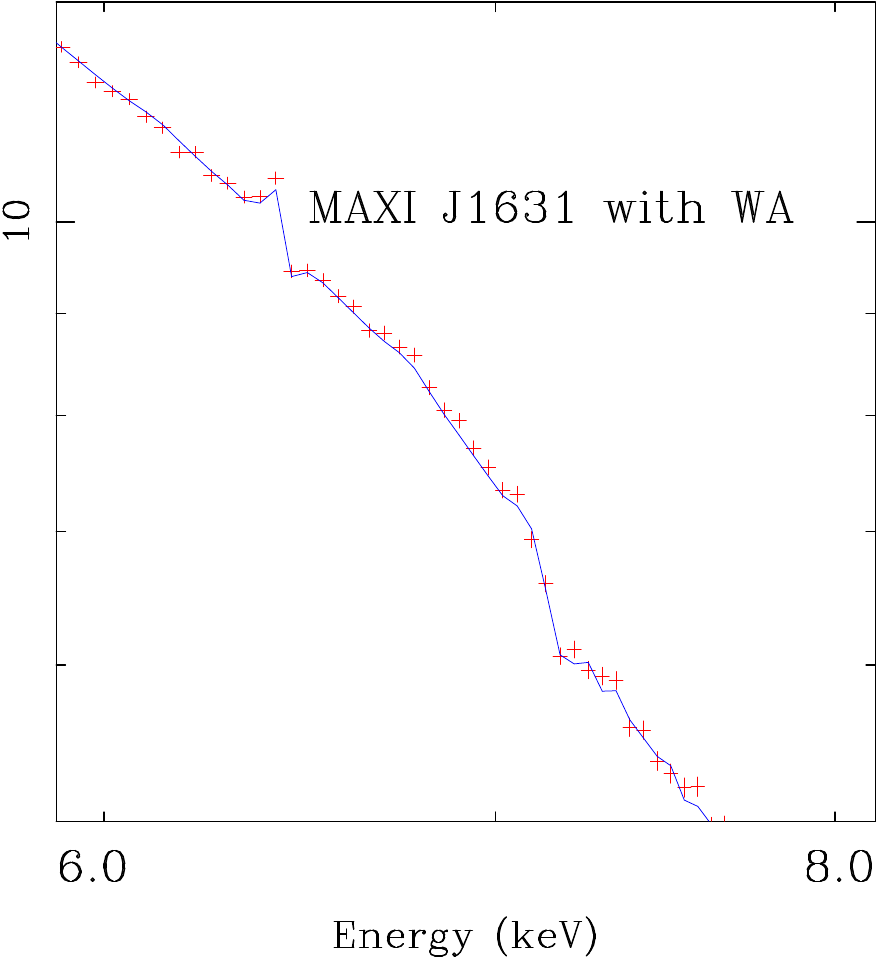} \\
\end{center}
\caption{The total model and the data plotted for the first part of MAXI J1631. The red points and blue curves correspond to the data and the theoretical best-fit model, respectively. The left plot shows the results for the data without a warm absorber (WA), and the right plot shows the results with a warm absorber modeled using {\tt XSTAR}}
ALT text: Effect of Warm absorber on MAXI J1631 \label{fig:wa}
\end{figure*}

\begin{figure*}
\begin{center}
\includegraphics[scale=0.4]{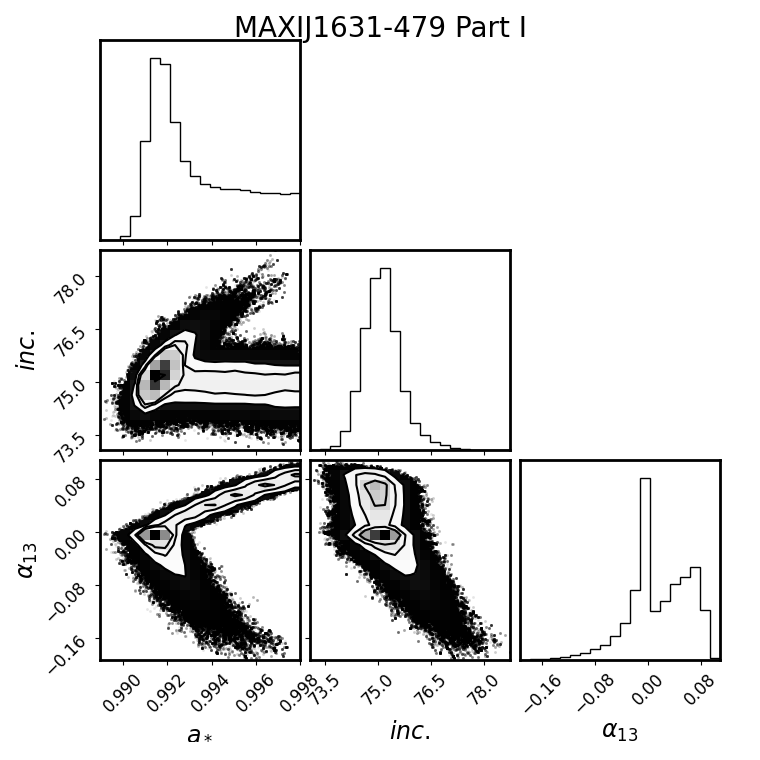}\includegraphics[scale=0.4]{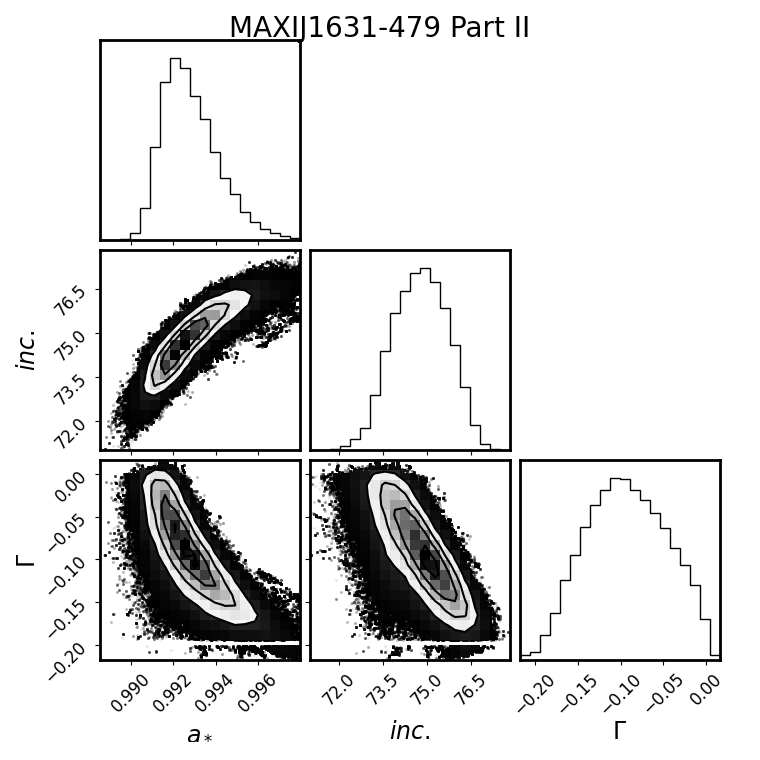}
\end{center}
\caption{Corner plot for the free parameter-pairs (excluding calibration constants and component normalization) for part I and part II observation of MAXIJ 1631-479 after the MCMC run. The 2D plots report the 1-, 2-, and 3-$\sigma$ confidence contours.\\
ALT text: MCMC plot for MAXI J1631 part I and part II.\label{fig:mcmcmaxi12}}
\end{figure*}

\subsection{MAXI J1631-479}

MAXI J1631-479 is a black hole discovered in 2018 when the source underwent an outburst and is situated very close (8.9’) to another X-ray transient AX J1631.9–4752. \textsl{NuSTAR} observed this source on 17, 27, and 29 January 2019 for the duration of 16.3, 10.1, and 14.4 ks, respectively \citep{Xu:2020vil}. In this case, the source region is 200” centered on the source. The corresponding background region is the annulus region of the inner radius and the outer radius to be 200” and 300” respectively. The \textsl{NuSTAR}  data are grouped to have a signal-to-noise ratio of 20 counts per bin. 

In Observation 1, the source is found to be in the disk dominant state, where the inner edge of the accretion disk is found to be at the innermost stable circular orbit (ISCO), which is assumed in the calculations involved in {\tt relxill\_nk}. The source is found to be in the power law dominant state in Observations 2 and 3. The disk is truncated, and the inner edge is not at the ISCO which violates the assumption that the inner edge of the disk is at ISCO \footnote{Disk truncation in the hard state has also been explored in numerical simulations \citep{2022ApJ...935L...1L, 2022MNRAS.517.5032D}.} \citep{1997ApJ...489..865E, Xu:2020vil, 2021ApJ...913...79T}. Hence, observations 2 and 3 cannot be analyzed with relxill\_nk, and only observation 1 is analyzed further. Significant spectral variability is found during this observation. Fig.~\ref{fig:maxi3} shows the \textsl{NuSTAR} light curve in the energy range of 3.0-79.0 keV, with part \Romannum{1} and part \Romannum{2} plotted in red and blue, respectively. The flux is increased by the factor of $\approx$3 during this period. We divided this observation into two parts based on flux variation to properly account for the variation in spectral analysis. The flux remained constant during the first part (part \Romannum{1}) and varied during the second part (part \Romannum{2}). 

Initially, the spectra were fitted with the combination of disk blackbody emission and power law affected by neutral absorption to inspect the reflection features present in the observation. In XSPEC, the phenomenological model is written as {\tt tbabs*(diskbb+powerlaw)}. In Fig.~\ref{fig:abspow}, we can see a broad and asymmetric emission line around 6 keV, which corresponds to the iron line and the Compton hump ($\sim$20 keV). In this case, thermal emission dominates up to 10 keV, which essentially provides the high-energy photons required for ionizing the Fe K-shell electron and producing Fe K$\alpha$ emission. Therefore, thermal emission is believed to play a major role in Fe fluorescence and in the increase of observed iron line flux, which is not possible with coronal illumination alone. We add {\tt relxillCp\_nk} model to account for relativistic reflection. We analyze both parts of the observation separately. \\

After the initial fit, there is some visible absorption around 7 keV, which is modeled using the {\tt XSTAR} model. It essentially models the warm absorber into several zones present around the black hole, and each zone has a characteristic column density $n_H$ and ionization parameter $\xi$. We freeze the iron abundance at the solar value. In XSPEC, the model can be written as 
\begin{equation}
{\tt tbnew*XSTAR*(diskbb + relxillCp\_nk)}
\end{equation}

Fig.~\ref{ratio}  shows the part \Romannum{1} and part \Romannum{2} of the observation fitted with the best-fit model. Table~\ref{tab-fit} reports the uncertainties for each parameter using MCMC simulation. For part \Romannum{1}, the reduced chi-square is found to be 1.27, which is high compared to other analyses, because the quality of the data at higher energies is of low quality, as seen in the phenomenological fit shown in Fig.~2. Addition of XSTAR fits the absorption edges around 6.7 and 7.1 keV and also improves the chi-square by $\sim$50 as shown in Fig.~\ref{fig:wa}. We freeze the inner emissivity at 10 and the outer emissivity at 3. The spin and inclination angle are found to be consistent with those found in previous studies ~\citep{Xu:2020vil}. For part \Romannum{2}, the reduced chi-square is 1.32, and we use simple power, which assumes the same emissivity index throughout the disk. The spin and inclination angle are consistent with both previous studies and the part \Romannum{1}. The emissivity profile of part \Romannum{1} observation is well fitted by the broken power law, whereas for part \Romannum{2}, a single power law describes the emissivity. For part \Romannum{1}, a broken power law, with an inner emissivity index of 10, would accumulate more emission in the inner region of the black hole, causing the reflection from the disk to increase. After the break radius of $\sim$7, the emissivity index becomes 3, which is the Newtonian profile. However, for part \Romannum{2}, the single power law with inner emissivity of $\approx$10 results in an increase in ionization of the relativistic reflection and the warm absorber around it. The warm absorber {\tt xstar} of part I is found to have higher column density and ionization compared to part \Romannum{1}. However, similar values of the inner temperature of the disk, electron temperature, ionization, and the photon index indicate that the change in dust or external absorbers causes the difference in flux between the two parts of observation 1. Fig.~\ref{fig:mcmcmaxi12} shows the corner plot of key parameters for both parts \Romannum{1} and part \Romannum{2}. The deformation parameter $\alpha_{13}$ includes the Kerr hypothesis with 99.73\% confidence for both parts and is also consistent with each other within 1~$\sigma$ errors. The errors associated with spin $a_*$ and the deformation parameter $\alpha_{13}$ for part \Romannum{1} and part \Romannum{1} is found to be 
\begin{equation}
 a_* = 0.993^{+0.003}_{-0.002}, \hspace{1cm} \alpha_{13} = 0.012^{+0.056}_{-0.034}
\end{equation}

and 

\begin{equation}
 a_* = 0.993^{+0.002}_{-0.001}, \hspace{1cm} \alpha_{13} = -0.094^{+0.052}_{-0.049}
\end{equation}
respectively. These constraints on spin and $\alpha_{13}$ obtained from the analysis of both parts are consistent with each other. The $Ref_{frac}$ is found to be more than unity for both parts of the observation, making the spectra power law dominated. However, a higher value of $Ref_{frac}$ is also possible due to the complex nature of the spectra. The corner plot for both parts shows a strong degeneracy of $Ref_{frac}$ with other parameters, which also affects the correct estimation of the parameter. The histogram of $\alpha_{13}$ in part \Romannum{1} observation shows the bimodal distribution. This may be due to two different regions of solution for this parameter, possibly due to variability in flux. Another reason for this bimodality is statistical limitation, and we will need more simulations to clarify the origin of such behavior, which is beyond the scope of this work. The $\chi^2$ improves by 1 and 5 for part I and part II, respectively, when the best-fit model is fit with $\alpha_{13}\neq 0$ as compared to that with $\alpha_{13} = 0$. The smaller difference in the chi-square for part I corresponds to the fact that the deformation parameter is closer to 0 compared to part II. 


\begin{figure*}
\begin{center}
\includegraphics[scale=0.8, height=40mm, width=120mm]{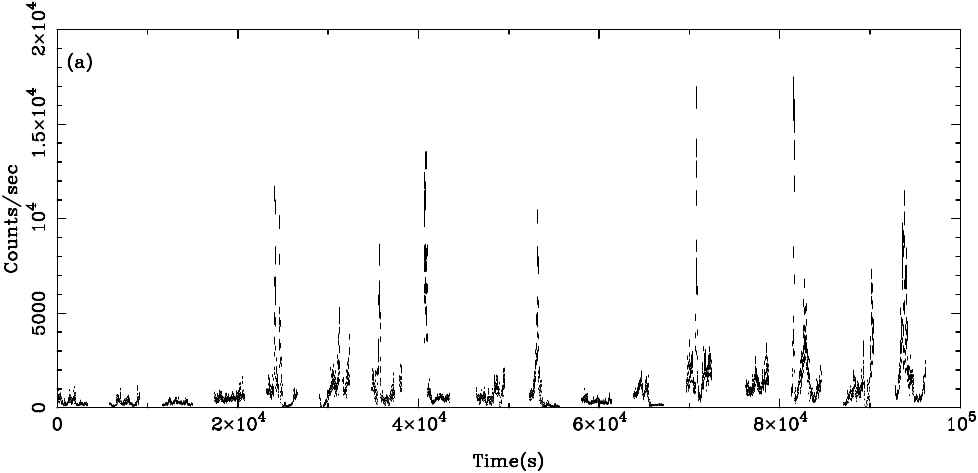}\\
\includegraphics[scale=0.8,height=40mm, width=120mm]{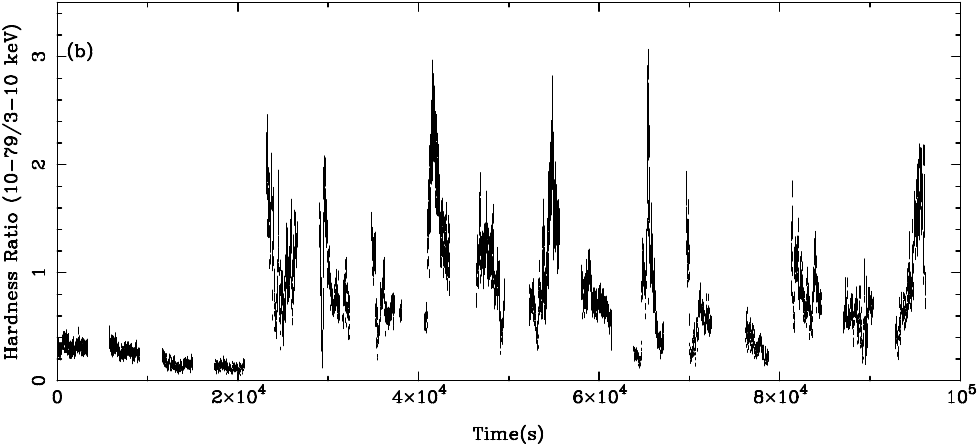} \\
\includegraphics[scale=0.4]{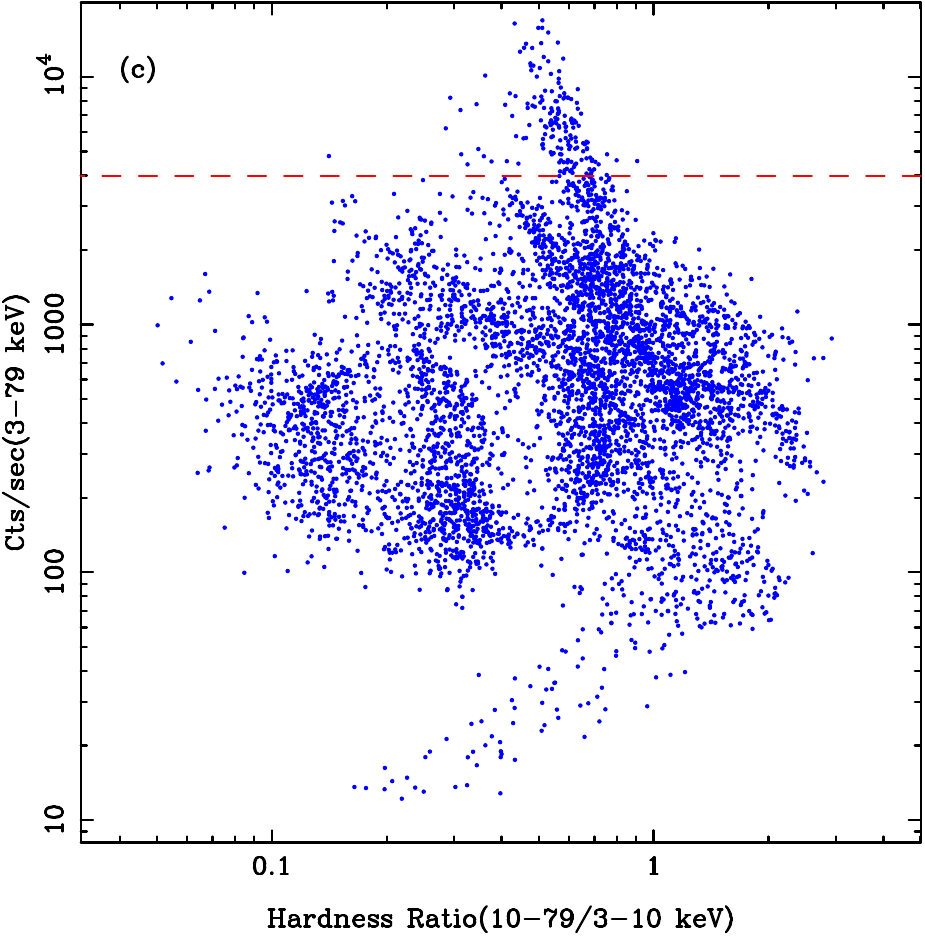}
\includegraphics[scale=0.4]{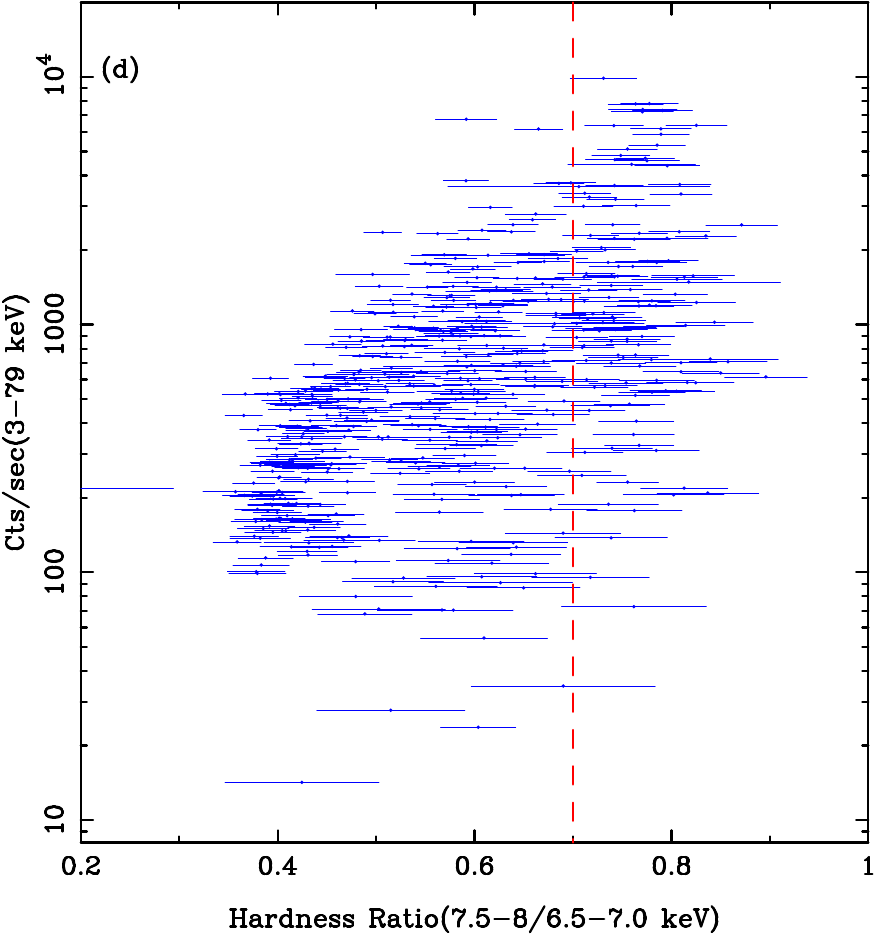} \\

\end{center}
\caption{{\it (a)} : the light curve of V404 analyzed in this work. {\it (b)}: hardness ratio between soft (3.0-10.0 keV) and 
hard (10.0-79.0 keV) X-ray waveband. {\it (c)}: hardness intensity diagram of the source during the observation. The horizontal red line signifies the flux threshold (4000 cts/s) for obtaining the flare spectra.{\it (d)}: The hardness intensity diagram is the same as the previous one, but for the hardness ratio between the energies in the range 7.5-8.0 keV and 6.5-7.0 keV.\\
ALT text: results for V404.}\label{fig:v404}
\end{figure*}

\begin{figure*}
\begin{center}
\hspace{-2cm}\includegraphics[scale=0.4]{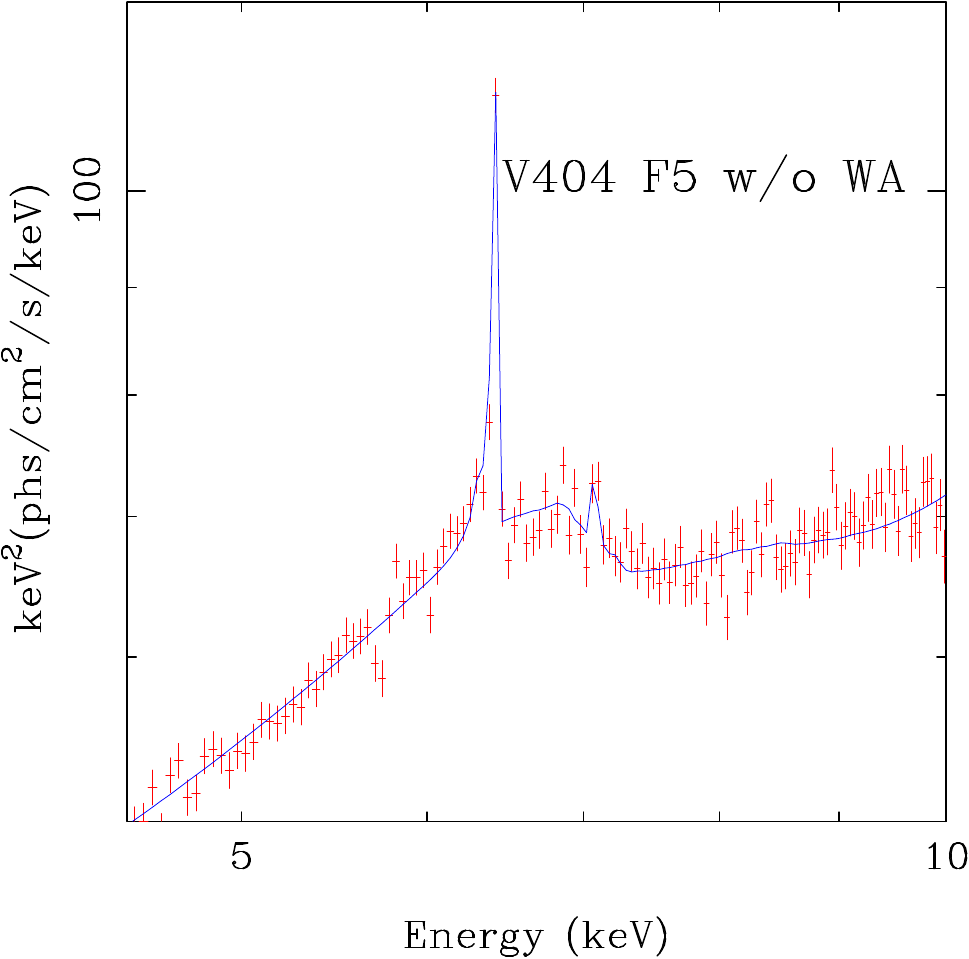}\includegraphics[scale=0.4]{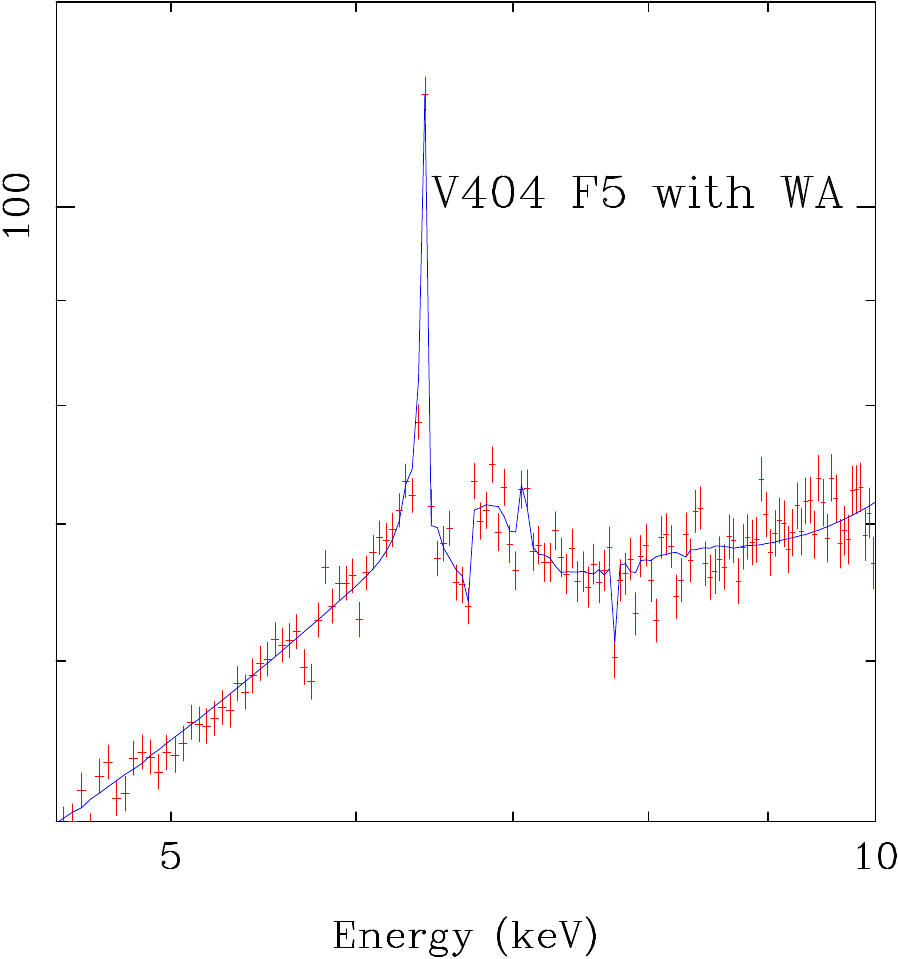} \\
\end{center}
\caption{The total model and the data plotted for the fifth flux state of V404. The red points and blue curves correspond to the data and the theoretical best-fit model, respectively. The left plot shows the results for the data without a warm absorber (WA), and the right plot shows the results with a warm absorber modeled using {\tt XSTAR}}
ALT text: Effect of Warm absorber on v404 \label{fig:wa1}
\end{figure*}

\begin{table*}
\centering
{\renewcommand{\arraystretch}{1.1}
\begin{tabular}{lcccc}
\hline\hline
Parameters& T15 & T37 & \multicolumn{2}{c}{MAXI J1631-479}\\
\hline
&&&Part I&Part II\\
\hline\hline
{\tt tbabs} \\
$N_{\rm H}$ [$10^{22}$~cm$^{-2}$] & $5.94^{+0.21}_{-0.17}$ & $12.51^{+0.29}_{-0.28}$ &  $6.10^{+0.67}_{-0.75}$ & $4.33^{+0.64}_{-0.62}$\\
\hline 
{\tt xstar} \\
$N_{\rm H}$ [$10^{22}$~cm$^{-2}$] &-&- & $1.77^{+0.99}_{-0.89}$  & $4.15^{+0.88}_{-0.82}$\\
$\log\xi$ [erg~cm~s$^{-1}$] & - & - & $2.04^{+0.08}_{-0.12}$ & $2.15^{+0.04}_{-0.05}$\\
\hline
{\tt diskbb} \\
$T_{in}$ & - & - & $1.016^{+0.003}_{-0.003}$ & $1.023^{+0.005}_{-0.005}$ \\
Norm & -& -& $5284^{+117}_{-144}$ & $4472^{+84}_{-139}$ \\
\hline
{\tt relxillCp\_nk} \\
$q_{\rm in}$ &$10^*$&    $10^*$ & $10^*$ & $9.81^{+0.14}_{-0.28}$\\
$q_{\rm out}$ & $4.53^{+0.84}_{-0.77}$ & $2.44^{+0.05}_{-0.05}$ & $3^*$ &-\\
$R_{\rm br}$ [$r_{\rm g}$] & $1.99^{+0.12}_{-0.12}$&$3.00^{+0.08}_{-0.08}$ & $6.84^{+0.23}_{-0.21}$&-\\
$a_*$ &$0.997^{+0.001}_{-0.001}$&$0.997^{}_{-0.001}$& $0.993^{+0.003}_{-0.002}$ & $0.993^{+0.002}_{-0.001}$\\
$i$ [deg] &$72.7^{+0.8}_{-0.8}$& $>6$& $75.1^{+0.5}_{-0.5}$ & $74.8^{+0.9}_{-0.9}$\\
$\Gamma$ & $2.05^{+0.01}_{-0.01}$ & $1.55^{+0.01}_{-0.01}$ & $3.39^{+0.01}_{-0.01}$ & $3.23^{+0.03}_{-0.03}$ \\
$\log\xi$ [erg~cm~s$^{-1}$] & $3.66^{+0.03}_{-0.07}$ & $3.28^{+0.02}_{-0.03}$& $2.65^{+0.06}_{-0.08}$ & $3.61^{+0.09}_{-0.09}$  \\
$A_{\rm Fe}$  & $1^*$ &$10^*$ & $1^*$& $1^*$ \\
$R_{\rm f}$ & $0.56^{+0.04}_{-0.04}$ &$0.31^{+0.04}_{-0.06}$ & $5.99^{+0.58}_{-0.54}$&$1.38^{+1.04}_{-0.99}$\\
$k_{Te}$ & $366^{+24}_{-46}$& $41^{+3}_{-4}$  & $198^{+133}_{-104}$ & $261^{+37}_{-75}$\\
$\alpha_{13}$ &$-0.016^{+0.013}_{-0.026}$&$-0.212^{+0.077}_{-0.095}$& $0.012^{+0.056}_{-0.034}$ &$-0.094^{+0.052}_{-0.049}$\\
$\alpha_{13}$ (3$\sigma$) &$-0.016^{+0.032}_{-0.109}$& $-0.212^{+0.214}_{-0.264}$& $0.012^{+0.085}_{-0.168}$ & $-0.094^{+0.102}_{-0.111}$\\
Norm  & $1.45^{+0.03}_{-0.11}$($10^{-3}$) & $6.19^{+0.24}_{-0.39}$($10^{-3}$) & $0.16^{+0.05}_{-0.07}$ & $0.53^{+0.02}_{-0.02}$\\
\hline
$\chi^2$/d.o.f. & 1569.10/1397 & 2591.04/2436 & 594.25/467 & 887.18/671\\
$\chi^2_\nu$& =1.12 & =1.06& 1.27 & 1.32 \\

\hline\hline
\end{tabular}
}
\caption{\rm Summary of the best-fit values of T15, T37, and MAXI J1631-479 . \label{tab-fit}}
\end{table*}

\subsection{V404 Cyg}

V404 Cyg (V404 afterward) is one of the closest black holes ($\sim$2.4 Kpc away) present in a binary system \citep{2009MNRAS.394.1440M, 2016A&A...587A..61C} with a K-type stellar companion and has a mass of 9-15M$\odot$ \citep{1996ApJ...460..437S, 2010ApJ...716.1105K}. It is a low-mass X-ray binary that experienced rare outbursts during which it became the brightest X-ray binary in the sky \citep{2014MNRAS.439.2771B, 2016ApJ...821..103R}. The first outburst since 1989 \citep{1999MNRAS.305..231Z} from the source was observed in the summer of 2015, which was followed by various multi-wavelength campaigns \citep{2015ApJ...813L..37K, 2015ApJ...813L..21N,2015A&A...581L...9R,2016MNRAS.459..554G, 2016ApJ...826...37J}. \textsl{NuSTAR} observed the source five times after the outburst. In this work, we focus on the first observation, which is split into two observation IDs (90102007002 and 90102007003) \citep{Walton:2016fso}. We need to turn off the filtering of a few hot pixels to take into account very high count rates and their rapid variability present throughout the observation. In addition to using the latest CALDB files, which have a database of such hot/flickering pixels, we also used the expression “STATUS=b0000xx00xx0xx000”, which controls the filtering. If we do not use such filtering and use the standard NUPIPELINE routine, then it is possible that some source photons during the flaring process could be eliminated. This filtering will keep these flare photons by not identifying them as hot/flickering. The source region of 160’’ with the source as the center is taken on a detector. Due to the very bright nature of the source, it is not possible to extract the background region from the same detector. As FPM consists of 4 detectors, the background region is taken from the other detector as far as possible from the source. As the data of this source comes from two consecutive observation IDs (90102007002 and 90102007003), we combine the FPMA spectra from these two observations using the ADDASCASPEC tool, and the same process is repeated to generate combined FPMB spectra. The spectra are binned in such a way that each bin would contain at least 50 counts per bin for the applicability of the $\chi^2$ statistics.

Fig.~\ref {fig:v404}(a) shows the light curve extracted from the \textsl{NuSTAR} observation following the method described in \citet{Walton:2016fso}. The extreme variability can be seen throughout the observation, with the count rate exceeding 20,000 counts/sec during the peak flares. The flux observed from the source increased by an order of magnitude very rapidly. The observation also displays spectral variability. Fig.~\ref {fig:v404}(b) shows the progression of the hardness ratio, which is the ratio of flux in the hard band (10-79 keV) and the soft band (3-10 keV). It shows that the hardness ratio remains almost constant for the first few orbits and then starts to vary rapidly, which coincides with the onset of flares in the observation. In Fig.~\ref {fig:v404}(c), we plotted the hardness intensity diagram for this observation, which is essentially the hardness ratio plotted against the flux in the whole allowed energy band (3-79 keV). Spectral states are not specifically defined, and no correlation is found in the majority of observations similar to \citet{Walton:2016fso}. The positive correlation can only be seen at lower count rates of up to 100 cts$^{-1}$, which is much lower compared to the counts during flares ($10^4$ cts$^{-1}$).

\begin{table*}
\centering
{\renewcommand{\arraystretch}{1.1}
\begin{tabular}{lccccc}
\hline\hline
Parameters& F1 & F2 & F3 & F4 &  F5\\
\hline\hline
{\tt tbabs} \\
$N_{\rm H}$ [$10^{22}$~cm$^{-2}$] & \multicolumn{5}{c}{$1^*$} \\
\hline 
{\tt xstar} \\
$N_{\rm H}$ [$10^{22}$~cm$^{-2}$] & $3.57^{+0.17}_{-0.37}$ & $4.27^{+0.39}_{-0.21}$ & $1.00^{+0.26}_{-0.82}$ & $1.26^{+0.41}_{-0.22}$ & $0.98^{+0.26}_{-0.16}$ \\
$\log\xi$ [erg~cm~s$^{-1}$] & $1.41^{+0.07}_{-0.11}$ & $2.61^{+0.07}_{-0.05}$ & $3.52^{+0.12}_{-0.14}$ &$3.54^{+0.13}_{-0.13}$ & $3.43^{+0.22}_{-0.19}$\\
{\tt relxillCp\_nk} \\
$q_{\rm in}$ & $8.4^{+1.1}_{-1.0}$ & $9.6^{+0.3}_{-0.6}$ & $9.4^{+0.5}_{-1.1}$ &$9.4^{+0.5}_{-0.9}$& $7.5^{+0.8}_{-0.6}$\\
$q_{\rm out}$ & $2.5^{+0.1}_{-0.1}$ & $2.38^{+0.05}_{-0.05}$ & $2.41^{+0.04}_{-0.04}$ & $2.43^{+0.03}_{-0.04}$& $2.46^{+0.05}_{-0.05}$\\
$R_{\rm br}$ [$r_{\rm g}$] & $1.6^{+0.4}_{-0.4}$ &$2.3^{+0.1}_{-0.1}$ & $2.08^{+0.11}_{-0.14}$ & $2.30^{+0.11}_{-0.10}$& $2.96^{+0.25}_{-0.22}$\\
$a_*$ & \multicolumn{5}{|c|}{$0.996^{+0.001}_{-0.004}$} \\
$i$ [deg] &  \multicolumn{5}{|c|}{$28^{+1}_{-1}$} \\
$\Gamma$ & $1.511^{+0.005}_{-0.004}$ & $1.560^{+0.005}_{-0.005}$ & $1.502^{+0.006}_{-0.006}$ & $1.525^{+0.006}_{-0.006}$ & $1.545^{+0.005}_{-0.007}$ \\
$k T_{\rm e}$ [keV] & $<225$ &$<395$& $39^{+1}_{-2}$ & $32^{+2}_{-1}$&$34^{+2}_{-1}$ \\
$\log\xi$ [erg~cm~s$^{-1}$] & $2.96^{+0.03}_{-0.07}$ & $3.04^{+0.01}_{-0.01}$ & $3.08^{+0.01}_{-0.02}$ & $3.13^{+0.02}_{-0.02}$ & $3.33^{+0.02}_{-0.01}$ \\
$A_{\rm Fe}$  &  \multicolumn{5}{|c|}{ $1.50^{+0.06}_{-0.06}$} \\
$R_{\rm f}$ & $0.23^{+0.02}_{-0.02}$ &$0.78^{+0.06}_{-0.06}$ & $0.65^{+0.06}_{-0.05}$ & $0.89^{+0.08}_{-0.08}$ & $1.17^{+0.12}_{-0.11}$ \\
$\alpha_{13}$ & \multicolumn{5}{|c|}{$-0.032^{+0.025}_{-0.051}$ }\\
$\alpha_{13}$ (3$\sigma$) & \multicolumn{5}{|c|} {$-0.032^{+0.037}_{-0.228}$}\\
Norm({\tt relxillCp\_nk}) & $0.071^{+0.002}_{-0.003}$ &$0.088^{+0.002}_{-0.002}$ & $0.128^{+0.005}_{-0.009}$ & $0.186^{+0.012}_{-0.021}$ & $0.318^{+0.043}_{-0.05}$ \\
Norm({\tt XillverCp}) & $0.036^{+0.003}_{-0.002}$ &$0.097^{+0.002}_{-0.004}$ & $0.074^{+0.007}_{-0.004}$ & $0.097^{+0.015}_{-0.016}$ & $0.197^{+0.025}_{-0.011}$ \\
\hline
$\chi^2_\nu$ &  \multicolumn{5}{|c|}{11540.3//10839 = 1.0608 }\\

\hline\hline
\end{tabular}
}
\caption{\rm Summary of the best-fit values of the model best fit for the observation of V404 analyzed in this work. \label{tab-fit1}}
\end{table*}

To check the presence of absorption during flares and to highlight the reflection features present in the observation, we obtained spectra for which the count rate exceeded 4,000 counts/sec. The total exposure is only about 1.4 ks, and the reflection features are clearly visible when fitted with the absorbed power law, as shown in figure~\ref{fig:abspow}. There is also an absorption edge around 7.5 keV, implying that additional absorption is present along with the Galactic one. There is no visible absorption in the flare spectrum. This implies that the flares also originate from the inner region of the accretion disk, and the absorption is unlikely to have any significant effect on very high count rates. 

To account for the rapid variability, we perform flux-resolved spectroscopy for this observation. We divide the data into five flux bins. To account for the variable absorption throughout the observation, we select only periods with low absorption on the basis of the flare spectrum. To explain these low-absorption periods, we define a quantity named the narrow band hardness ratio ($R_{edge}$), which is essentially the ratio between the softer band (6.5-7.0 keV) and the harder band (7.5-8.0 keV). These bands are chosen such that the softer band is just below the sharp absorption edge observed in the average spectrum, and the harder band is just above the sharp edge. This scheme is used to assess the strength of the absorption throughout the observation. In these narrow bands, strong absorption corresponds to the soft spectrum and a lower hardness ratio. Therefore, we need to include the period that has $R_{edge}$ higher than a certain threshold, which is taken as 0.7. Fig.~\ref{fig:v404}(d) shows the hardness intensity diagram for the hardness defined in the energy range 7.5-8.0 keV and 6.5-7.0 keV. The threshold is denoted by a red dashed line. 

We select the data after imposing constraints on flux bins and $R_{edge}$ i.e., we select the data with $R_{edge}$ $>$ 0.7 throughout the observation with count rates less than 4,000 cts/sec and then resolve the data in four flux bins: 100-500, 500-1000, 1000-2000, and 2000-4000 cts$^{-1}$. For the flux bin of more than 4000, we took the whole period as it already has $R_{edge}$ $>$ 0.7. There are five flux states in total, named as F1-5. The lower limit of the data used in the analysis is 100 cts/sec because, below this, the flux in the broadband is correlated with the hardness ratio. 

Now, we have five flux states with low absorption. We analyzed all five flux states simultaneously. After fitting the data with a relativistic reflection model (which includes power-law emission), an absorption can be seen at around 7.5 keV. We added the non-relativistic reflection model {\tt xillver} to explain this absorption and other features in the data, which could be the result of distant reflection, possibly from a cloud. Adding {\tt xillver} to the model removes the corresponding residuals and also improves $\chi^2$ by $\sim$1053. Finally, the model used to describe this black hole system consists of a power-law emission from the corona, a reflection spectrum from the disk (both nearby and away from the black hole), galactic absorption, and the intrinsic absorption of the source. In XSPEC, the model is written as :

\vspace{0.1cm}
{\tt tbabs*XSTAR*(relxillCp\_nk + xillverCp)}
\vspace{0.1cm}

The column density of galactic absorption is kept frozen at $10^{22}$ cm$^{-3}$ \citep{Walton:2016fso} and is the same for all flux states. The column density and ionization of {\tt XSTAR} is kept free among different flux states. {\tt relxillCp\_nk} models the Comptonized relativistic reflection model, which comes from the innermost region of the black hole. The broken emissivity profile is used and kept free among different flux states, as it is believed that the difference in the flux could be due to different emission rates. Spin ($a_*$), inclination ($i$), and iron abundance ($A_{Fe}$) are tied between various flux levels. The ionization parameter $log\xi$ is kept untied between the flux levels because the accretion disk is believed to be ionized differently and produces different flux states. {\tt xillverCp} models the distant reflector, which is essentially not affected by the strong gravity of the central compact object. The incident spectrum that produces both distant non-relativistic and relativistic reflection spectra is assumed to be produced by the same mechanism, and hence the photon index, electron temperature of both {\tt relxillCp\_nk} and {\tt xillverCp} are tied. As {\tt relxillCp\_nk} includes the continuum, we only include the reflected component from {\tt xillverCp}, which is done
by setting the reflection fraction to -1. The accretion disk is assumed to be "cold" (not ionized enough) at larger distances, and hence the ionization parameter is frozen to 0. The iron abundance is tied between the {\tt relxillCp\_nk} and {\tt xillverCp}.

Fig~\ref{ratio}  shows the ratio of the data fitted to the best model for all five flux states for V404. Tab.~\ref{tab-fit1} reports the uncertainties on the parameters that are variable during the fit. We obtained a good fit for all five flux states, and the combined $\chi^2$ is found to be 1.06. The residuals remain after fitting the best model decrease as we go from flux F1 to F5. The column density of the warm absorber is lower for high flux states, which implies that the absorption affects the spectrum less as the flux increases. There is an absorption edge around 7.5 keV that is fitted with the XSTAR model. In addition, there are some features around 6.7 keV that are fitted with a warm absorber as shown in Fig.~\ref{fig:wa1}. The broken emissivity profile used here was found to have a very steep inner emissivity ($\sim$10) and a low outer emissivity ($\sim$3.0) with a break radius in the range 1.5-3.0.  The source is found to have a large spin and a small inclination angle, which is consistent with previous studies of the source \citep{Walton:2016fso}. Fig.~\ref{fig:mcmcv404} shows the triangle plot for the posterior distribution of globally defined parameters for the best-fit model. $\alpha_{13}$ recovers the Kerr solution with 99.73\% confidence. The 1~$\sigma$ uncertainty associated with spin $a_*$ and deformation parameters $\alpha_{13}$ for V404 is given by

\begin{equation}
 a_* = 0.996^{+0.001}_{-0.004}, \hspace{1cm} \alpha_{13} = -0.032^{+0.025}_{-0.051}   
\end{equation}

The $\chi^2$ for the non-Kerr model improves almost negligibly as compared to that for the model when $\alpha_{13} = 0$, which is possibly due to the value of $\alpha_{13}$ being very close to the Kerr solution.

\begin{figure*}
\begin{center}
\includegraphics[scale=0.5]{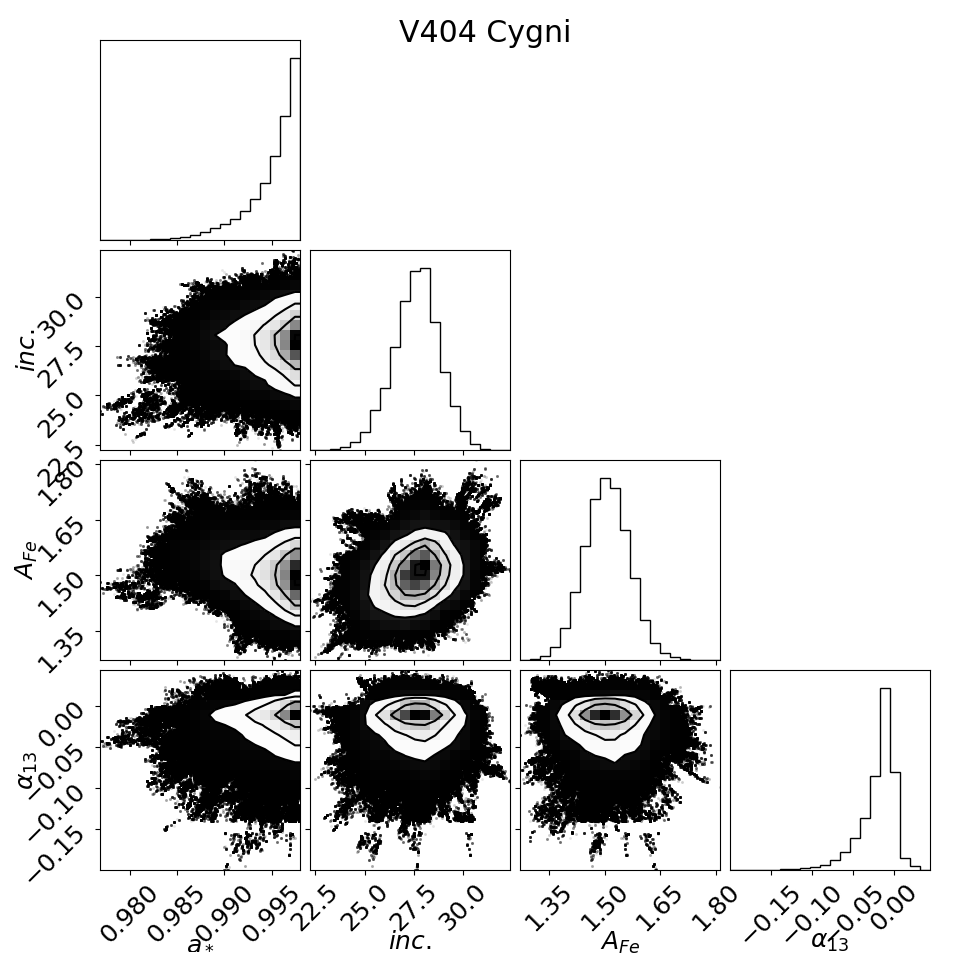}
\end{center}
\vspace{-0.5cm}
\caption{Corner plot for the free parameter-pairs (excluding calibration constants and component normalization) for V404 Cygni after the MCMC run. The 2D plots report the 1-, 2-, and 3-$\sigma$ confidence contours.\\
ALT text: MCMC plot for V404. \label{fig:mcmcv404}}
\end{figure*}

\begin{figure*}
    \centering
    \includegraphics[scale=0.7]{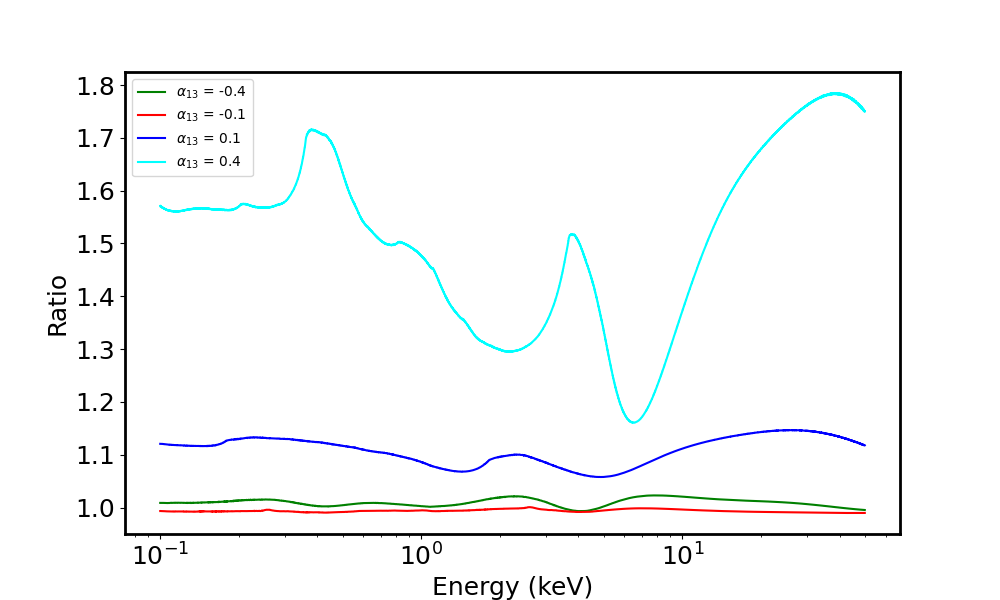}
    \caption{Plot illustrating the dependence of the reflection spectrum on the space-time metric of the black hole. The curves of different colors correspond to the ratio of the reflection spectrum for different values of the deformation parameter ($\alpha_{13}$) to the reflection spectrum for the Kerr black hole ($\alpha_{13}$ = 0).
    ALT text: Reflection spectrum for different values of deformation parameters.\label{fig:alpha}}
\end{figure*}

\begin{figure*}
\centering
\includegraphics{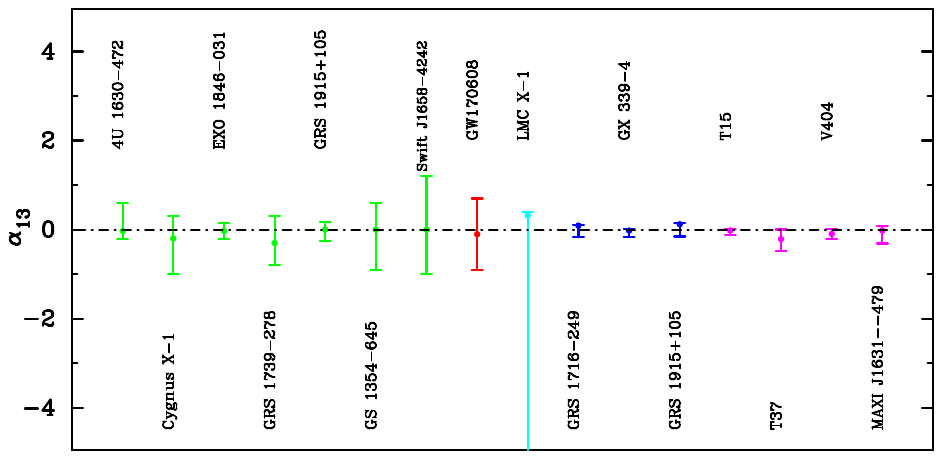}
\caption{The error associated with deformation parameter $\alpha_{13}$ obtained for various tests of general relativity. The magenta curves are the measurements obtained in this work. The red curves are for the measurements from gravitational wave events. The green curves correspond to the results using the iron line method. The cyan-colored measurement is estimated using the continuum fitting method. The blue curve shows the result obtained by using the continuum-fitting method and the iron line method simultaneously.}
ALT text: measurement of deformation parameter for different sources.\label{fig:12}
\end{figure*}


\section{Discussion and conclusions} \label{s:dis}

In this work, we have analyzed highly complicated X-ray \textsl{NuSTAR} observations of four X-ray binaries: T15, T37, MAXI J1631--479, and V404. We have used the non-Kerr extension of the state-of-the-art relativistic model to analyze these observations and obtain constraints compared to what was obtained in \citet{2021ApJ...913...79T}. Such comparable constraints can only be obtained through advanced reduction techniques applied to the observations, like flux-resolved and time-resolved spectroscopy. It is highly crucial to model the warm absorbers and absorption, if present, in the observation, which could be very specific in every case. For example, the flux changed by 15\% in a single observation for MAXI J1631--479. We divide the observation in such a way that the variation in each part remains constant and thus accounts for the variability. Even time-resolved spectroscopy does not produce good statistics, possibly due to the complex nature of the observation. This might be explained by using a complex spectral model, which is beyond the scope of this work. Hence, we used the standard relativistic model with warm absorbers to model the source. In V404, multiple flares are present in the observation, which are divided into various flux states to account for variability and absorption. In addition, we also need to impose $R_{edge}<$~0.7 to analyze the data that are least affected by absorption. We used the version of the galactic absorption model {\tt Tbnew} which deals with heavy absorption. Therefore, adopting such sophisticated data reduction schemes would prove useful in explaining the physical conditions around the black hole and obtaining the stringent constraints on metric and accretion disk parameters, leading to testing general relativity with such highly absorbed X-ray observations.


Fig.~\ref{fig:alpha} shows the ratio of the reflection spectrum for different values of the deformation to the reflection spectrum when it vanishes ($\alpha_{13}$=0). For simulating the spectra, the values of $a_*$ and $i$ are assumed to be 0.998 and 30\degree, respectively. The iron abundance is assumed to be solar and the logarithmic ionization parameter is fixed at 3. The emissivity profile is assumed to be a broken power law with the values of $q_{in}$, $q_{out}$, and $R_{br}$ fixed at 10.0, 3.0, and 3.0, respectively. We plotted two negative values (-0.4 and -0.1) and two positive values (0.1 and 0.4) of $\alpha_{13}$. For both negative values of $\alpha_{13}$, the ratio of the reflection spectrum is close to 1.0, which essentially means that they are similar to the reflection spectrum of a Kerr black hole. The reflection spectrum computed for $\alpha_{13}$~=~0.1 has 10\% higher flux, on average, as compared to that of the Kerr black hole. The flux of the reflected spectrum at $\alpha_{13}$=0.4 is 70\% high at 0.4 keV (80\% at 40 keV) compared to the flux for $\alpha_{13}$ = 0.0. It, therefore,  implies that the positive values of the deformation parameter adversely affect the amplitude of the reflection spectrum compared to its negative values.

Fig.~\ref{fig:12} shows the 3~$\sigma$ confidence interval on the measurement of $\alpha_{13}$ estimated by analyzing X-ray data from various X-ray binaries (this work and \citet{2021ApJ...913...79T}). We also show the measurement of $\alpha_{13}$ from the LIGO/Virgo data \citep{Cardenas-Avendano:2020xtw}. The gravitational event GW170608, shown in Fig.~\ref{fig:12}, provides the strongest constraints, which are comparable to the constraints obtained in this work. For each source, we first fit the data with $\alpha_{13} = 0$, and then the deformation parameter is allowed to vary freely. Each time, the chi-square improves, although less. As the value of the deformation parameter always includes the Kerr solution, the results of Kerr and the non-Kerr model will not differ much in terms of $\chi^2$, and the parameters are also found to be constrained. This could also be  a limitation of the resolution of the features provided by the existing instrument. Future telescopes could distinguish between these models and help us assess the true nature of a black hole.

The data analyzed in this work are initially fitted with the different models using the minimum $\chi^2$ statistics in {\tt XSPEC}. After getting the best-fit model, the MCMC simulations are performed with 500 walkers and 12,000 iterations, which is equivalent to six million simulations. The number of walkers is chosen so that the whole physical range of all parameters is explored simultaneously. Some parameters are fixed in the analysis because their variation is insensitive to fits and is unconstrained.  We fix it at a reasonable value, and the effect of such parameters on the fit is also checked with MCMC simulations, where the parameter is found to vary across the whole range. These simulations have been used to estimate errors in the best-fit parameters, which are purely statistical in nature and do not include systematic ones. This statistical error becomes very small as we have run six million iterations on the best-fit model for each source. In the case of part II of MAXI J1631, the spin is negatively correlated with the deformation parameter but has a negative correlation with the inclination, which is similar to the results from part I. Interestingly, for V404, the spin is found to have negligible correlation with other parameters. For T15 and T37, the spin is positively correlated with the inclination but negatively correlated with $\alpha_{13}$ for negative values. For positive values of $\alpha_13$, the correlation is positive for both sources, which is also true for the part I observation of MAXI J1631. Such correlations between spin and $\alpha_{13}$ could also be intrinsic to the model \citet{Abdikamalov:2019yrr}. Therefore, these correlations would underestimate the errors provided by the MCMC simulations, which should be understood with caution.

The impact of $\alpha_{13}$ on the reflection spectrum arises because of its effect on the ISCO radius, which strongly depends on the spin parameter.  \citet{Bambi:2015kza} shows that the spin and $\alpha_{13}$ are strongly anti correlated. In recent studies, it has been shown that if the spin is high ($a_*>0.95$) and could be tightly constrained with observation, then $\alpha_{13}$ could also be tightly constrained. For example, in \citet{Tripathi:2020qco}, measurement (90 \% confidence interval) of spin and $\alpha_{13}$ for LMC X-1 are $a_*<0.56$ and $-2.78<\alpha_{13}<0.36$, respectively, which are poorly constrained. On the other hand, the spin measurement for GX 339-4 is found to be $0.99<a_*<0.998$ for the best fit model, which is comparable to what has been found in this work\citep{Tripathi:2020dni}. The deformation parameter for GX 339-4, similar to spin, is tightly constrained, and the measurement is $-0.038<\alpha_{13}<-0.001$. In our work, the spin is found to be very tightly constrained. Due to degeneracy, the constraints on the deformation parameter are also very tight compared to what was found in \citet{Bambi:2015kza}. The simulations done in \citet{Bambi:2015kza} assume a spin value of 0.8, which is low compared to the spin values found in this work. Therefore, the constraints obtained on the deformation parameter are poor compared to those found in high-spin value fits because of this intrinsic degeneracy between $a_*$ and $\alpha_{13}$.


Several assumptions are made in the relativistic reflection models used in this work. These simplifications also introduce systematic uncertainties in the measurement of the deformation parameter. We will discuss these simplifications and their potential effects on the results. The disk emissivity profile could be very crucial in determining the appropriate reflection spectrum of a source. In this work, we tried the emissivity profiles included in {\tt relxill\_nk}: simple power law, broken power law, and  the lampost. We select the best profile that yields the minimum $\chi^2$. In {\tt relxill\_nk}, the inner edge of the accretion disk is assumed to be at the innermost stable circular orbit (ISCO). The assumption is valid for  sources in soft states having the Eddington ratio of ~$\approx$0.05--0.3 \citep{Steiner:2010kd, 2011MNRAS.414.1183K}. In this work, we assume that the inner edge of the accretion disk lies at the ISCO, even in the hard states \citep{Bambi:2016sac, 2021ApJ...913...79T}, which implies that the observations that show a truncated disk are not suitable for our analysis. We performed analysis for each observation with both lamppost and broken power-law emissivity profiles and chose the emissivity profile for the observation from the $\chi^2$ value. For example, broken power-law emissivity improves $\chi^2$ by $\sim$100 for V404 compared to the lamppost. {\tt relxill\_nk} assume the accretion disk to be infinitesimally thin. In reality, the disk has a finite thickness, which could increase with the increase in mass accretion rate. As the sources in this work are maximally rotating, the thickness of the disk has a negligible effect on the measurement of $\alpha_{13}$ as shown in \citet{Abdikamalov:2020oci}.


Current reflection models ignore the effects of radiation from the plunging region \citep{Cardenas-Avendano:2020xtw}. If the emission from within the plunging regions matters, sources with low spins would have significant emission from within the ISCO and produce spectra that would look like having been caused by high spins. However, recently, \citet{2025arXiv250702583S} showed using numerical simulations that the plunging region is likely to be highly ionized and therefore does not produce significant emission lines, causing a non-significant effect on the spin measurement. The {\tt relxill\_nk} version used in this work assumes a constant electron density $n_e$ = $10^{15}$ $cm^{-3}$, which is too low for the X-ray binaries with a high mass accretion rate. The sources analyzed in this work have high mass accretion rates, and therefore, this effect needs to be explored further. Studies by \citet{Zhang:2019ldz} and \citep{Tripathi:2020dni} conclude that the electron density has a negligible effect on the measurement of $\alpha_{13}$. {\tt relxill\_nk} does not take into account the effect of returning radiation. A fraction of the reflection radiation emitted from the accretion disk returns to the disk due to the strong light-bending effect in the innermost regions of a black hole. \citet{2023EPJC...83..838R} studied the effect of returning radiation on the reflection spectrum through simulations and found that the current versions of reflection models (without returning radiation) can be used to test the Kerr metric.


The analysis using relxill\_nk is more than merely fitting the data with relxill, having one extra parameter. It is physically motivated in the sense that the value of this extra parameter, which is $\alpha_{13}$, signifies the presence of a non-Kerr spacetime near the inner-most region of the black hole. At the very least, allowing for such a deformation can absorb potential systematic uncertainties in the astrophysical and spectral models. By introducing an additional degree of freedom beyond the standard models, we gain the ability to probe such systematics explicitly. The statistical uncertainty on the deformation provides a quantitative measure of how much the true spacetime or modeling uncertainties might deviate from theoretical predictions. The statistical uncertainty should become smaller with the future instruments like NewAthena \citep{2025NatAs...9...36C} with high-resolution spectroscopy, while any systematic biases in models might remain. Such analyses, therefore, allow us not only to constrain possible deviations but also to quantify systematic biases in reflection models.

In \citet{2021ApJ...913...79T}, simple sources are analyzed where the absorption is negligible, and therefore, the systematics are limited. The observations analyzed in this work have systematics due to heavy absorption, in addition to the usual systematics associated with the reflection model discussed previously. In this work, we reduce the systematics due to heavy absorption by sophisticated data reduction and analysis techniques and successfully constrain $\alpha_{13}$ between -1.0 and 0.5, which includes the Kerr solution. Getting consistent zero deformation values signifies that even with the complicated spectra, the models work well for the data quality of \textsl{NuSTAR}. \citet{Tripathi:2020dni} addresses the systematics due to the reflection model by analyzing data with different models, which is beyond the scope of this work. 

Measuring deviations from the Kerr metric is one way to falsify the underlying theory. To date, all observational results are consistent with Kerr within statistical errors. If we ever observe a statistically significant deviation, it may not simply mean GR is incorrect. Rather, it would open the possibility of exploring more intricate physics through improved theoretical models and observational techniques. Even then, demonstrating a deviation from Kerr would help us challenge or refine our understanding of the spacetime metric. The more such observations are found in X-ray and in other techniques, the stronger our conclusions about the spacetime geometry will become.

This paper attempts to address the systematics due to the reflection model and heavy absorption, and still demonstrates the capability of X-ray reflection spectroscopy to constrain the deviations from the Kerr metric. These four measurements open a new frontier, showing that absorbed sources can also be employed to test the Kerr hypothesis.  Although these results do not provide the existence of a non-Kerr black hole, they certainly strengthen the claim of these objects being described by the Kerr metric, which is also a confirmation of the assumption of explaining the space-time metric with the Kerr metric in most of the relativistic reflection models currently employed in X-ray astronomy. The future X-ray telescopes, like \textsl{eXTP} \citep{Zhang:2016ach} and \textsl{NewAthena} \citep{2025NatAs...9...36C}, would provide a higher energy resolution around the iron line. This improved sensitivity will prove helpful in obtaining stringent constraints on a non-Kerr metric.

\section*{Acknowledgments}

This work was supported by the Tianshan Talent Training Program (grant No.\ 2023TSYCCX0099), the CAS `Light of West China' Program (grant No. 2021-XBQNXZ-005), and the National SKA Program of China (grant No.\ 2022SKA0120102). A.T. acknowledges the support from the Xinjiang Tianchi Talent Program. This work was also supported by the Urumqi Nanshan Astronomy and Deep Space Exploration Observation and Research Station of Xinjiang (XJYWZ2303). SS acknowledges support from the Shanghai Super Postdoctoral fellowship. GM acknowledges the support from the China Scholarship Council (CSC), Grant No. 2020GXZ016647. 




\end{document}